\documentclass[a4paper]{cas-dc}

\usepackage[numbers]{natbib}
\usepackage[toc,acronym]{glossaries}
\usepackage{adjustbox}
\usepackage{tikz}   
\usepackage{pgfplots}
\pgfplotsset{compat=1.13} 
\usepackage{pgfplotstable}
   \definecolor{webgreen}{rgb}{0,.5,0}
\definecolor{webblue}{rgb}{0,0,.8}
\definecolor{webred}{rgb}{0.8, 0, 0}   
\definecolor{webbrown}{rgb}{.6,0,0}
\definecolor{webyellow}{rgb}{0.98,0.92,0.73}
\definecolor{webgray}{rgb}{.753,.753,.753}

\pgfplotscreateplotcyclelist{my color list}{%
	solid, color=webblue, every mark/.append style={solid, fill=webblue}, mark=*\\%
	densely dashdotted, color=webgreen, every mark/.append style={solid, fill=webred},mark=diamond*\\%
	densely dotted, color=webbrown, every mark/.append style={solid, fill=webgreen}, mark=triangle*\\%
	loosely dashed, color=webred, every mark/.append style={solid, fill=webbrown},mark=*\\%
	dotted, color=webblue, every mark/.append style={solid, fill=webyellow}, mark=square*\\%
	densely dotted, color=webgreen, every mark/.append style={solid, fill=gray}, mark=otimes*\\%
	dashed, color=webbrown, every mark/.append style={solid, fill=gray},mark=diamond*\\%
	densely dashed, every mark/.append style={solid, fill=gray},mark=square*\\%
	dashdotted, every mark/.append style={solid, fill=gray},mark=otimes*\\%
	dashdotdotted, every mark/.append style={solid},mark=star\\%
}

\def\tsc#1{\csdef{#1}{\textsc{\lowercase{#1}}\xspace}}
\tsc{WGM}
\tsc{QE}
\tsc{EP}
\tsc{PMS}
\tsc{BEC}
\tsc{DE}
\makeglossaries

\begin{document}
\let\WriteBookmarks\relax
\def\floatpagepagefraction{1}
\def\textpagefraction{.001}
\shorttitle{The roofline of parallelized performance gain}
\shortauthors{J. V\'egh}

\title [mode = title]{The performance wall of parallelized sequential computing:\\
	the dark performance and the roofline of performance gain}                      
\tnotemark[1]

\tnotetext[1]{This work is supported by the National
	Research, Development and Innovation Fund of
	Hungary, under  K funding scheme Projects no. 125547 and 132683, as well as ERC-ECAS support of project 861938 is acknowledged}

\author{J\'anos V\'egh}[
                        orcid=0000-0001-7511-2910
                        ]
\ead[url]{Vegh.Janos@gmail.com}

%
\address[1]{Kalim\'anos BT, Hungary. 4032 Debrecen, Koml\'ossy 26.}
%
%


%
%

\cortext[cor1]{Corresponding author}
%

\begin{abstract}
The computing performance today is developing mainly using parallelized sequential computing,
in many forms. 
The paper scrutinizes whether the performance of that type of computing has an upper limit.
The simple considerations point out that the theoretically possible upper bound is practically
achieved, and that the main obstacle of to step further is the presently used computing
paradigm and implementation technology.
In addition to the former "walls", also the "performance wall" must be considered.
As the paper points out, similarly to the "dark silicon", also the "dark performance" 
is always present in the parallelized many-processor systems. 
\end{abstract}

\begin{graphicalabstract}
	Using parallel computing for achieving the needed higher performance
	adds one more limitation to the already known ones.
	The performance gain of parallelized sequential processor systems
	has a theoretical upper limit, derived from the laws of nature, the paradigm
	and the technical implementation of computing. Based on the different limiting
	factors, the paper derives some theoretical upper limit of the performance gain. 
	From the rigorously controlled database of supercomputers,
	performance gains of the ever-built supercomputers are analyzed
	and compared to the theoretical bound.
	It is shown that the present implementations already achieved the theoretical upper bound.
	The upper bound of the performance gain for processor-based brain simulation is also derived.

\includegraphics{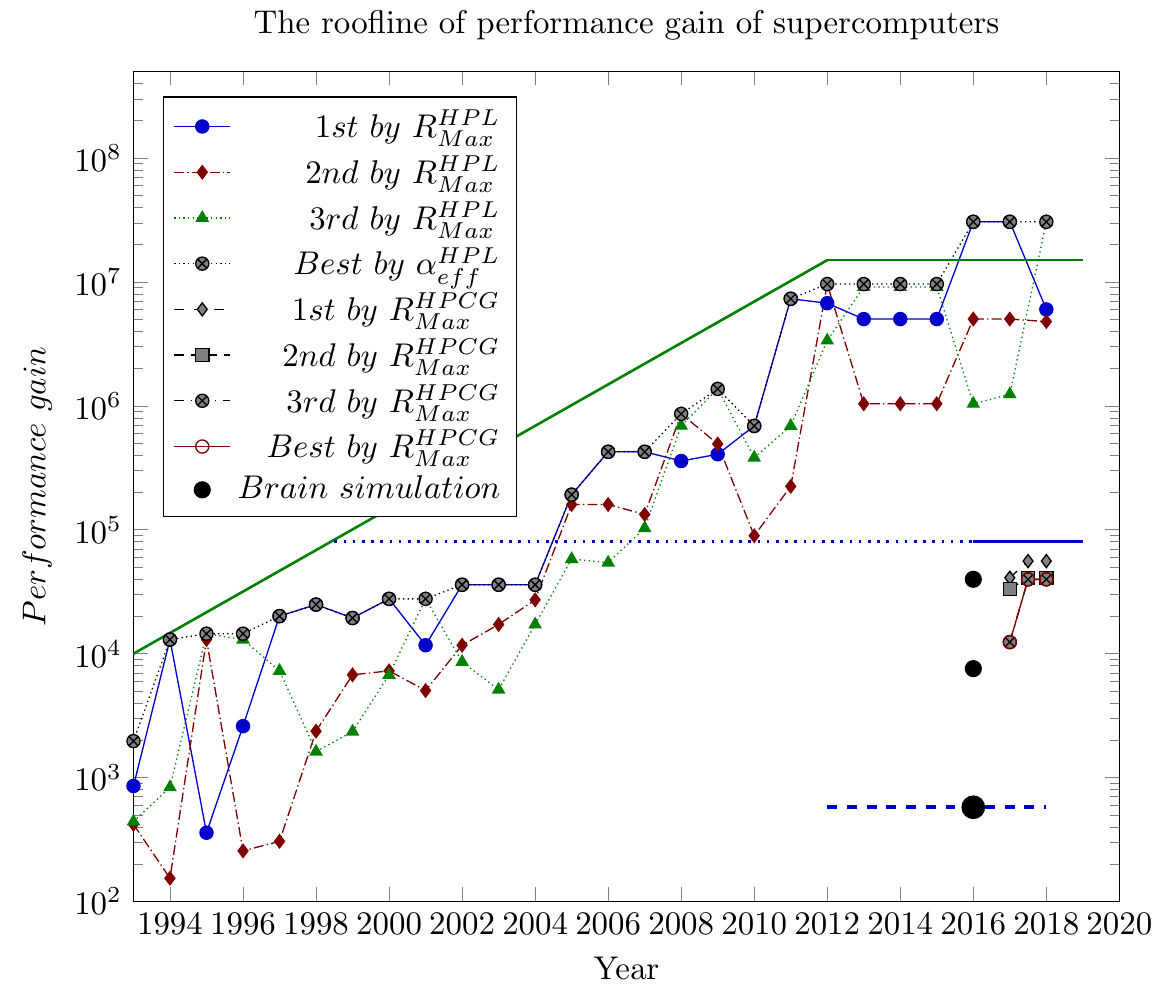}
\end{graphicalabstract}

\begin{highlights}
\item A new limitation of computing performance is derived
\item A new merit for parallelization is introduced
\item The theoretically possible performance gain is already achieved
\item The performance of brain simulation is compared with that of the benchmarks 
\end{highlights}

\begin{keywords}
efficiency\\
 single-processor\\ performance\\ performance gain\\ supercomputer\\ brain simulation
\end{keywords}

\newacronym{AI}{AI}{Artificial Intelligence}
\newacronym{ASIC}{ASIC}{Application Specific Integrated Circuit}
\newacronym{CMC}{CMC}{Configurable Architecture \gls{MC}}
\newacronym{CPS}{CPS}{Cyber-Physical System}
\newacronym{CPU}{CPU}{Central Processing Unit}
\newacronym{EMPA}{EMPA}{Explicitly Many-Processor Approach}
\newacronym{FPGA}{FPGA}{Field Programmable Gate Array}  
\newacronym{GPGPU}{GPGPU}{General-Purpose Graphics Processing Unit}
\newacronym{GPU}{GPU}{Graphics Processing Unit}
\newacronym{HW}{HW}{hardware}
\newacronym{HPC}{HPC}{High Performance Computing}
\newacronym{HPL}{HPL}{High Performance LINPACK~\cite{HPCG_List:2016}} 
\newacronym{HPCG}{HPCG}{High Performance Conjugate Gradients~\cite{HPCG_List:2016}}
\newacronym{ICB}{ICB}{Inter Core Block}
\newacronym{ILP}{ILP}{Instruction Level Parallelism}
\newacronym{ISA}{ISA}{Instruction Set Architecture}
\newacronym{I/O}{I/O}{Input/Output}
\newacronym{MC}{MC}{Multi-Core and/or Many-Core}
\newacronym{MLP}{MLP}{Memory Level Parallelism}
\newacronym{OoO}{OoO}{Out-of-Order}
\newacronym{OS}{OS}{operating system}
\newacronym{PU}{PU}{Processing Unit}
\newacronym{QT}{QT}{quasi-thread}
\newacronym{PC}{PC}{Program Counter}
\newacronym{PD}{PD}{Propagation delay}
\newacronym{RC}{RC}{reconfigurable}
\newacronym{RT}{RT}{real-time}  
\newacronym{SoC}{SoC}{System-on-Chip}
\newacronym{SPA}{SPA}{Single Processor Approach}
\newacronym{RMC}{RMC}{Reconfigurable Architecture \gls{MC}}
\newacronym{SV}{SV}{supervisor}
\newacronym{SW}{SW}{software}
\newacronym{TLP}{TLP}{Thread Level Parallelism}
\definecolor{darkmagenta}{rgb}{0.55, 0.0, 0.55}

\maketitle

\section{Introduction}

The parallelization in the computing today becomes more and more focus~\cite{HennesseyParallelCACM2009}.
After that the growth of the single-processor performance has stalled~\cite{GameOverYelick:2011},
the only hope for making computing systems with higher performance is to assemble them from a large number of sequentially working computers. "\textit{Computer architects have long sought the "The City of Gold" (El Dorado)
	of computer design: to create powerful computers simply by connecting many existing smaller ones.}"~\cite{RISCVarchitecture:2017}. 
Solving the satisfying scalability of the task, however, is not simple at all. One of the major motivations of the origin of the Gordon Bell Prize was to increase the resulting \textit{performance gain} of the parallelized sequential systems: "\textit{a speedup of at least
200 times on a real problem running on a general purpose
parallel processor}"~\cite{GordonBellPrize:2017}.

It is  known from
the very beginnings, that the "\textit{Validity of the Single Processor Approach to Achieving Large-Scale Computing Capabilities}"~\cite{AmdahlSingleProcessor67,ScalingParallel:1993} is at least questionable, and the really large-scale tasks
(like building supercomputers comprising millions of single processors or building brain simulators from single processors) may face serious scalability issues.
After the initial difficulties, for today
the large-scale supercomputers are stretched to the limits~\cite{StrechingSupercomputers:2017}, 
and although the EFlops (payload) performance has not yet been achieved, already the $10^4$ times higher
performance is planned~\cite{ChinaExascale:2018}.
Many of the planned large-scale supercomputers, however, are delayed, canceled, paused, withdrawn, re-targeted, etc. It looks like that
in the present "gold rush" it has not been scrutinized 
\textit{whether the resulting payload performance of parallelized systems has some upper bound}.
In the known feasibility studies  this aspect remains out of sight in USA~\cite{NSA-DOE_HPC_Report_2016}, in  EU~\cite{EUActionPlan:2016}, 
in Japan~\cite{JapanExascale:2018} and in China~\cite{ChinaExascale:2018}. The designers did not follow "\textit{that system designers must make the effort to understand the relevant characteristics of the
benchmark applications they use, if they are to arrive at
the correct design decisions for building larger multiprocessor systems}"~\cite{ScalingParallel:1993}.
The rules of game are different for the segregated processors and the parallelized ones~\cite{VeghModernParadigm:2019};
the larger are the systems, the more remarkable deviations appear in their behavior.
In addition to the already known limitations~\cite{LimitsOfLimits2014}, valid for segregated processors, a new limitation, valid for parallelized processors appears~\cite{VeghPerformanceWall:2019}.

This paper presents that the well-known Amdahl's law,
the commonly used computing paradigm (the single-processor approach~\cite{AmdahlSingleProcessor67}) and the commonly used implementation technology together 
form a strong upper bound for the performance  of the parallelized sequential computing systems, and the experienced failures, saturation, etc. experiences can be attributed to approaching and attempting to exceed that upper bound. In section~\ref{sec:amdahlclassic}
\textit{"We realize that Amdahl's Law is one of the few, fundamental laws of computing"}~\cite{AmdalsLaw-Paul2007},
and reinterpret it for the targeted goal.
The section introduces the mathematical formalism used and derives the deployed logical merits.
Based on the idea of Amdahl, in section \ref{sec:ourmodel} a general model of parallel computing in constructed.  
In this (by intention) strongly simplified model the contributions are classified either as parallelizable or nonparallelizable ones.
The section provides an overview of the components contributing to the non-parallelizable fraction of processing, and attempts to reveal their origin and behavior. By properly interpreting the contributions, it  shows that under different conditions different contributions can have the role of being the performance-limiting factor.
The section interprets also the role of the benchmarks, and explains why the different benchmarks produce different results.
Section~\ref{sec:InherentLimit} shows that the parallelized sequential computing systems have their inherent performance bound, and shows numerous potential limiting factors. 
The section also provides some supercomputer-specific numerical examples.

The last two sections directly underpin the previous theoretical discussion with measured data. In section~\ref{sec:SuperCaseStudies} some specific supercomputer features are discussed, where the case is well documented and the special case enables to draw general conclusions.
This section also discusses the near (predictable) future of large-scale supercomputers and their behavior.
Section~\ref{sec:statistical} draws some statistical conclusions, based on the available supercomputer database, containing rigorously verified, reliable data for the complete supercomputer history.
The large number of data enables, among others, to derive conclusions about a new law of the electronic development, to tell where and when it is advantageous to apply graphic accelerator units, as well as to derive a "roofline" model of supercomputing.

\section{Amdahl's classic analysis}\label{sec:amdahlclassic}

\subsection{Origin and interpretation}
The most commonly known and cited limitation on parallelization speedup (\cite{AmdahlSingleProcessor67}, the so called \emph{Amdahl's law}) 
considers the fact that some parts ($P_i$) of a code can be parallelized,
some ($S_i$) must remain sequential.
Amdahl only wanted to draw the attention to that when putting together
several single processors, and using \gls{SPA}, the available speed gain due to using large-scale computing capabilities \textit{has} a theoretical upper bound.
He also mentioned that data housekeeping (non-payload calculations)
causes some overhead, 
and  that \emph{the nature of that overhead appears to be sequential, independently of its origin}.
\index{Amdahl's law} \index{parallelization speedup}

\subsection{Validity}\label{sec:AmdahlValidity}
{A general misconception (introduced by successors of Amdahl) is to assume that Amdahl's law
	is valid for software only and that the non-parallelizable fraction means
	something like the ratio of numbers of the corresponding instructions.
	Amdahl in his famous paper speaks about
	"\textit{the  fraction  of  the  computational  load}"
	and explicitly mentions, in the same sentence and same rank, algorithmic reasons like 
	"\textit{computations  required   may  be  dependent  on  the  states  of  
		the   variables   at   each   point}"; architectural aspects like "\textit{may  be strongly   dependent   on  sweeping  through  the  array  
		along  different   axes  on  succeeding  passes}"
	as well as
	"\textit{physical problems}" like 
	"\textit{propagation   rates   of  different  physical  effects may be quite different}".
	His point of view is valid also today: 
	one has to consider \textit{the load of the complex \gls{HW}/\gls{SW} system,
		rather than some segregated component}, and his idea describes parallelization imperfectness of any kind.
	When applied to a particular case, however, one shall scrutinize which contributions can actually be neglected.
	
	Actually,\textit{ Amdahl's law is valid for any partly parallelizable activity (including computer unrelated ones) and the non-parallelizable fragment  shall be given as the ratio of the time spent with non-parallelizable activity 
		to the total time}. The concept was frequently successfully utilized on quite unexpected fields, as well as misunderstood and abused (see~\cite{UsesAbusesAmdahl:2001,Gustafson:1988}).}

As discussed in~\cite{LimitsOfLimits2014}
\begin{itemize}
	\item many parallel computations today
	are \textit{limited by several forms of communication and synchronization}
	\item the parallel and sequential runtime components are only slightly affected
	by cache operation
	\item  the \textit{wires get increasingly slower relative to gates}
\end{itemize}

\subsection{Amdahl's case under realistic conditions}
\label{sec:AmdahlRealistic}

The realistic case  is  that
parallelized parts are \emph{not} of equal length (even if they comprise exactly the same 
instructions). The hardware operation in modern processors may execute them in 
considerably different times; for examples see \cite{Vegh:2014:ICSOFTsemaphore,HennessyArchitecture2007}, and references cited therein;
operation of hardware accelerators inside a core, or network operation between processors, etc.
One can also see  that the time required to control parallelization is not negligible and varying;
 representing another source of performance bound.

The static correspondence between program chunks and
processing units can be very inefficient: all assigned processing units must wait the delayed unit.
The measurable performance does not match
the nominal performance:
leading to the appearance of the "\textit{dark performance}":
the processors cannot be utilized at the same time, much similar to how
fraction of cores (because of energy dissipation)  cannot be utilized at the same time, 
leading to the issue "\textit{dark silicon}"~\cite{DarkSilicon2012,Esmaeilzadeh:2015:AAP:2830689.2830693}.

Also, some capacity is lost if the number of computing resources exceeds number of parallelized chunks.
If number of processing units is smaller than that of the parallelized threads,
severals "rounds" for the remaining  threads must be organized,
with all disadvantages of duty of synchronization~\cite{YavitsMulticoreAmdahl2014,SynchronizationEverything2013}.
In such cases it is not possible to apply Amdahl's Law directly: the actual architecture is too complex or not known. 
However, in all cases the speedup can be measured and 
expressed in function of the number of processors.

\subsubsection{Factors affecting parallelism}

Usually,  Amdahl's law is expressed as 

\begin{equation}
S^{-1}=(1-\alpha) +\alpha/k \label{eq:AmdahlBase}
\end{equation}

\noindent where $k$ is the number of parallelized code fragments, 
$\alpha$ is the ratio of the parallelizable fraction to the total,
$S$ is the measurable speedup. 
The assumption can be visualized that (assuming many processors)
in $\alpha$ fraction of running time processors are processing data,
in (1-$\alpha$) fraction they are waiting (all but one). That is $\alpha$ describes
how much, in average, processors are utilized.
Having those data, the resulting speedup can be estimated. 

For the today's complex systems, to calculate $\alpha$ is hopeless, but
for a system under test, where  $\alpha$ is not \textit{a priory} known,
one can derive from the measurable speedup  $S$ 
an \emph{effective parallelization} factor as

\begin{equation}
\alpha_{eff} = \frac{k}{k-1}\frac{S-1}{S} \label{equ:alphaeff}
\end{equation}
\index{parallelism!effective}
\index{effective parallelism}

\noindent Obviously, this is not more than $\alpha$ expressed in terms of $S$ and $k$ from Equ.~(\ref{eq:AmdahlBase}).
So, for the classical case, $\alpha = \alpha_{eff}$; which simply means that
in \emph{ideal} case the actually measurable effective parallelization 
achieves the theoretically possible one.
In other words, $\alpha$ describes a system the \emph{architecture} of which is completely known,
while $\alpha_{eff}$ characterizes a system the \emph{performance} of which is known from experiments.
Again in other words,  $\alpha$ is the \emph{theoretical upper limit}, which can hardly be achieved,
while $\alpha_{eff}$ is the \emph{experimental actual value}, that describes the complex architecture and the actual conditions. 

The value of $\alpha_{eff}$ can then be used to refer back to Amdahl's classical
assumption even in realistic cases when the detailed architecture is not known.
On one side, the speedup $S$  can be measured and $\alpha_{eff}$ can be utilized
to characterize measurement setup and conditions~\cite{Vegh:2017:AlphaEff}, 
how much from the theoretically possible maximum parallelization is realized.
On the other side, the theoretically achievable 
$S$ or $\alpha_{eff}$ can be guessed from some general assumptions.

In the case of real tasks a Sequential/Parallel Execution Model \cite{YavitsMulticoreAmdahl2014} shall be applied, which cannot use 
the simple picture reflected by $\alpha$, but $\alpha_{eff}$ gives a good merit
of the degree of parallelization for the duration of  executing the process on the given hardware configuration,
and can be compared to the results of technology-dependent parametrized formulas\footnote{
	Just notice here that passing parameters among cores as well as blocking each other (of course, through the \gls{OS}) are all a kind of synchronization or communication, and their amount differs task by task.}.
Numerically ($1-\alpha_{eff}$) equals to $f$ value, established theoretically~\cite{Karp:parallelperformance1990}.
\index{parallelism!supercomputer}

To the scaling of parallel systems several models
can be applied, and all they can be goof on their
specific field. However, one must recall that
"The truth is
that\textit{ there is probably no specific parameter scaling principle that can be universally applied~\cite{ScalingParallel:1993}}."
This also means, that the validity of the scaling methods for extremely large number of processors must be scrutinized.

\subsection{Efficiency of parallelization}\label{sec:ParallelEfficiency}
The distinguished constituent in Amdahl's classic analysis is the parallelizable payload fraction $\alpha$,
all the rest (including wait time, communication, system contribution and any other non-payload activities) goes into the apparently "sequential-only" fraction  according to this extremely simple model.

When using several processors, one of them makes the sequential-only calculation, the others are waiting\footnote{In a different technology age, the same phenomenon was already described: "\textit{Amdahl argued that most parallel programs have
	some portion of their execution that is inherently serial
	and must be executed by a single processor while others
	remain idle}."~\cite{ScalingParallel:1993}}
(use the same amount of time).
In the age of Amdahl the number of processors was small
and the contribution of the \gls{SW} was high,
relative to the contribution of the parallized \gls{HW}. Because of this,
the contribution of the \gls{SW} dominated the sequential part,
so the value of $(1-\alpha)$ really could be considered as a constant.
The technical development resulted in decreasing all non-parallelizable components; the large systems today can be idle also because of \gls{HW} reasons.

Anyhow, when calculating speedup, one calculates
\begin{equation}
S=\frac{(1-\alpha)+\alpha}{(1-\alpha)+\alpha/k} =\frac{k}{k(1-\alpha)+\alpha}
\end{equation}
hence  \textit{efficiency}\footnote{This quantity is almost exclusively used to describe computing performance of multi-processor systems.
	In the case of supercomputers, $\frac{R_{Max}}{R_{Peak}}$ is provided, which is identical with $E$} (how speedup scales with number of processors)
\begin{equation}
E = \frac{S}{k}=\frac{1}{k(1-\alpha)+\alpha}\label{eq:soverk}
\end{equation}

\begin{figure}
	\maxsizebox{1.1\columnwidth}{!}
	{
			\includegraphics[width=.8\textwidth]{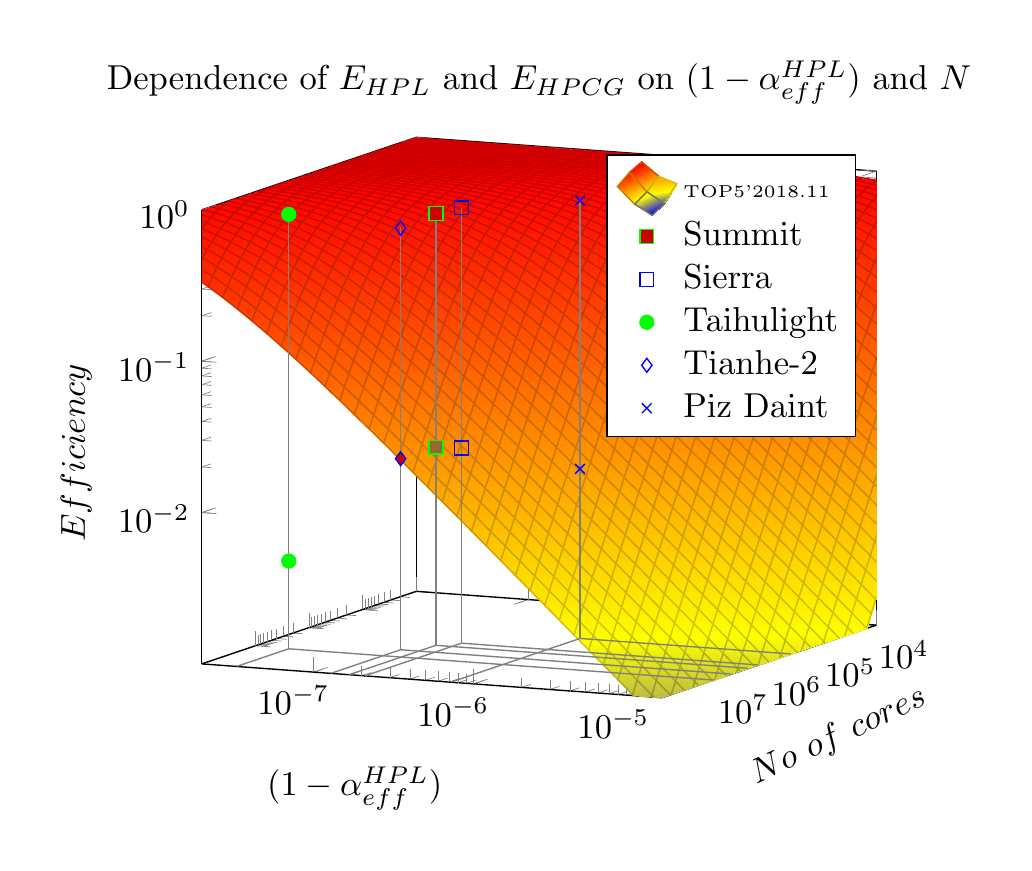}
		}
	\caption{The  dependence of computing efficiency of parallelized sequential computing systems on the parallelization efficacy and the number of cores. The surface is described by Eq.~(\ref{eq:soverk}), the data points for the TOP5 supercomputers are calculated from the publicly available database~\cite{Top500:2016}
}
\label{fig:EffDependence2018Log}
\end{figure}

This means that according to Amdahl, as presented in Fig.~\ref{fig:EffDependence2018Log},  the \textit{efficiency depends both on the
	total number of processors in the system and on the perfectness}\footnote{As it will explained below, in a higher-order approximation the value of $(1-\alpha)$ itself also depends on the number of the processors.} of the parallelization. 
The perfectness comprises two factors: 
the theoretical limitation and the engineering ingenuity.

If parallelization is well-organized (load balanced, small overhead, right number of \gls{PU}s), $\alpha$  saturates at unity (in other words: sequential-only fraction approaches zero),
so the tendencies can be better displayed through using $(1-\alpha_{eff})$ in the diagrams below.

The importance of this practical term $\alpha_{eff}$ is underlined by that
it can be interpreted and utilized in many different areas~\cite{Vegh:2017:AlphaEff,VeghBrainAmdahl:2019} and 
the achievable speedup (the maximum achievable performance gain when using infinitely large number of processors) can easily be derived from Equ.~(\ref{eq:AmdahlBase}) as

\begin{equation}
G=\frac{1}{(1-\alpha_{eff})} \label{eq:AmdahlMax}
\end{equation}

Provided that the value of $\alpha_{eff}$ does not depend on the number of the processors, for a homogenous system the total payload performance is

\begin{equation}
P_{total~payload}=G\cdot P_{single~processor} \label{eq:PayloadMax}
\end{equation}

\noindent i.e. \textit{the total payload performance can be increased
	by increasing the performance gain or by increasing the single-processor performance, or both}.
Notice, however, that increasing the single-processor performance
through accelerators also has its drawbacks and limitations~\cite{Vegh:StatisticalConsiderations:2017},
and that the performance gain and the single-processor performance are players of the same rank in defining the payload performance.

\subsubsection{Connecting efficiency and $\alpha$}

Through using Equ.~(\ref{eq:soverk}), $\frac{S}{k}$ can be equally good for describing efficiency of parallelization of a setup,
but anyhow a second parameter, the number of the processors $k$ is also required. From Equ.~(\ref{eq:soverk})
\begin{equation}
\alpha_{E,k} = \frac{E k -1}{E (k-1)}\label{eq:alphafromr}
\end{equation}

\noindent This quantity depends on both $E$ and $k$, but in some cases
it can be assumed  that $\alpha$ is independent
from the number of processors. This seems to be confirmed by data calculated from several publications
as was noticed early,

At this point one can notice that $\frac{1}{E}$ in Equ.~(\ref{eq:soverk}) is a linear function of number of processors, and its slope equals to
$(1-\alpha_{eff})$. The value calculated in this way is denoted by $(1-\alpha_\Delta)$.
Its numerical value is quite near  to the value  calculated  (see Equ.(~\ref{equ:alphaeff})) 
using all processors,
and so it is not displayed in the rest of figures. 
This also means that from efficiency data
one can estimate value of $\alpha_\Delta$ even for intermediate regions,
i.e. without knowing the execution time on a single processor
(from technical reasons, it is the usual case for supercomputers).
From a handful of processors one can find out if the 
supercomputer under construction can have hopes to beat\footnote{At least a lower bound on $\alpha_{eff}$ (i.e. a higher bound on parallelization gain) can be derived.} the No. 1 in Top500~\cite{Top500:2016}.
This result can also be used for investment protection.


\subsubsection{Time to organize parallelization}

The timing analysis given above can be applied to different kinds of parallelizations,
from processor-level parallelization (instruction or data level parallelization, in nanoseconds range)
to \gls{OS}-level parallelization (including thread-level parallelization using several processors or cores, in microseconds range),
to network-level  (between networked computers, like grids, in milliseconds range).
The principles are the same \cite{SynchronizationEverything2013},
independently of the kind of implementation.
In agreement with \cite{YavitsMulticoreAmdahl2014},
housekeeping overhead is always present (and mainly depends on the \gls{HW}+\gls{SW} architectural solution) and remains a key question.
The main focus is always on to reduce its effect.
Notice that the application itself also comprises some (variable) amount of sequential contribution.

The actual speedup (or effective parallelization) depends strongly on the 'tricks' used during implementation.
Although \gls{HW} and \gls{SW} parallelisms are interpreted differently~\cite{HwangParallelism:2016}, they even can be combined
\cite{ChandyParallelism:2009}, resulting in hybrid architectures. 
For those greatly different architectural solutions it is  hard even to interpret  $\alpha$,
while $\alpha_{eff}$ enables to compare
different implementations (or the same implementation under different conditions).

\section{Our model of parallel execution}\label{sec:ourmodel}

As mentioned in section~\ref{sec:AmdahlValidity}, Amdahl listed different reasons why losses in the "computational load"
can occur. 
To understand the operation of computing systems working in parallel, one needs to extend Amdahl's original (rather than that of the successors') model in such a way,
that the non-parallelizable (i.e. apparently sequential) part comprises contributions from \gls{HW}, \gls{OS}, \gls{SW} and \gls{PD}, and also some access time is needed  for reaching the parallelized system.
The technical implementations of the different parallelization methods show up infinite variety, so here a (by intention) strongly simplified model is presented.
Amdahl's idea enables to put everything\footnote{Although the modern diagnostic methods enable to separate the theoretically different contributions and measure their value separately, unfortunately their effect is summed up in the non-parallelizable fraction. The summing rule is not simple: some sequential contribution may occur in parallel with some other, like the propagation delay with the \gls{OS} functionality; and it must also be considered whether the given non-parallelizable item contributes to the main thread or to one of the fellow threads.}
that cannot be parallelized into the sequential-only fraction.
The model is general enough to discuss qualitatively some examples of parallely working systems, neglecting different contributions
as possible in the different cases. The model can also be converted to a limited validity  technical (quantitative) one.

\subsection{Formal introduction of the model}\label{sec:ParallelizationModel}

\begin{figure}
	\includegraphics{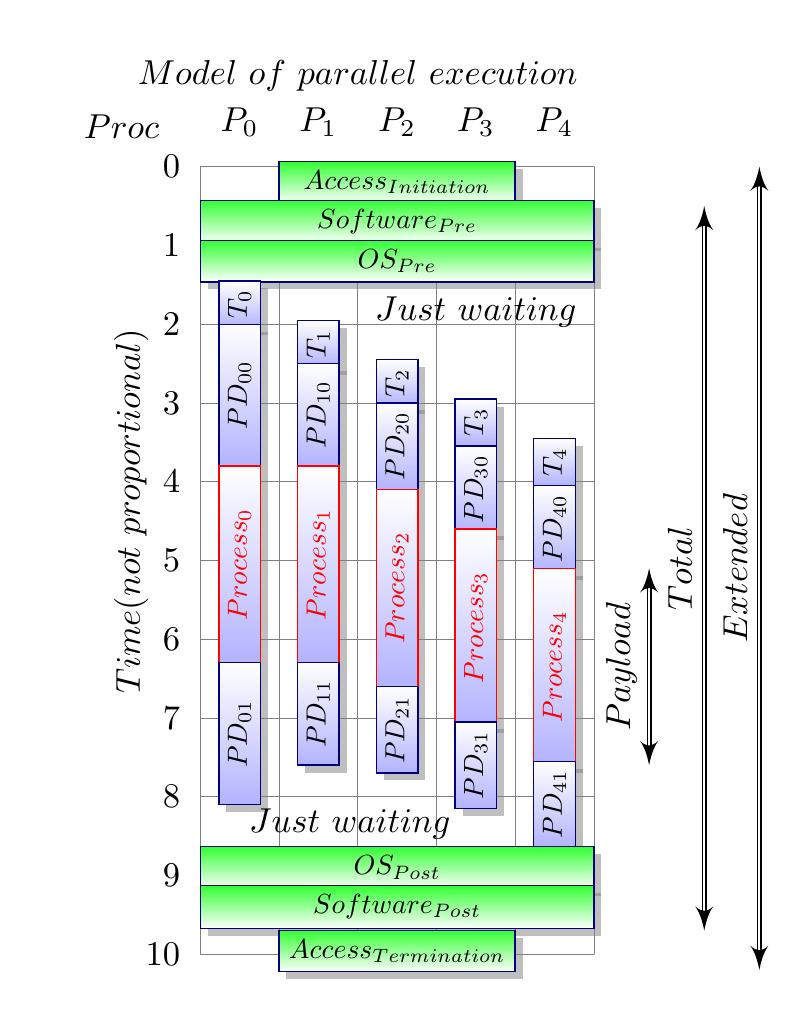}
	\caption{
		The extended Amdahl's model of parallelizing sequential processing (somewhat idealistic and not proportional)
	}
	\label{fig:Ourmodel}
\end{figure}

The contributions of the model component $XXX$ to $\alpha_{eff}$ 
will be denoted by $\alpha_{eff}^{XXX}$ in the following.
Notice the different nature of those contributions.
They have only one common feature: \textit{they all consume time}.
The extended Amdahl's model is shown in Fig.~\ref{fig:Ourmodel}.
The  vertical scale displays the actual activity for processing units shown on the horizontal scale. 

Notice that our model assumes no interaction between processes
running on the parallelized systems in addition to the absolutely necessary minimum: starting and terminating the otherwise independent processes, which take parameters at the beginning and return results at the end.
It can, however, be trivially extended to the more general case when processes must share some resource (like a database, which shall provide different records for the different processes), 
either implicitly or explicitly. Concurrent objects have inherent sequentiality~\cite{InherentSequentiality:2012}, and synchronization and
communication among those objects 
considerably increase~\cite{YavitsMulticoreAmdahl2014} the non-parallelizable fraction
(i.e. contribution $(1-\alpha_{eff}^{SW})$), so in the case of extremely 
large number of processors special attention must be devoted to their role on efficiency of the application on the parallelized system.

Let us notice that all contributions have a role
during measurement: the effect of contributions due to \gls{SW}, \gls{HW}, ~\gls{OS} and  \gls{PD} cannot be separated,
though dedicated measurements can reveal their role, at least approximately.
The relative weights of the different contributions are very different for the different parallelized systems,
and even within those cases depend on many specific factors,
so in every single parallelization case a careful analysis is required.

\subsubsection{Access time}\label{sec:AccessTime}
Initiating and terminating the parallel processing is usually made from within the same computer,
except when one can only access the parallelized computer system from another computer (like in the case of clouds).
This latter access time is independent from the parallelized system,
and \textit{one must properly correct for the access time when derives timing
	data for the parallelized system}. Amdahl's law is valid only for properly selected computing system.
This is a one-time, and usually fixed size time contribution.

\subsubsection{Execution time}

The execution time \textit{Total} covers all processings on the parallelized system.
All applications, running on a parallelized system, must make some non-parallelizable activity at least before beginning and after terminating parallelizable activity. 
This \gls{SW} activity represents what was assumed by Amdahl as the total sequential fraction\footnote{Although some \gls{OS} activity was surely included, Amdahl assumed some 20~\% \gls{SW} fraction, so the other contributions
	could be neglected compared to \gls{SW} contribution.}.
As shown in Fig.~\ref{fig:Ourmodel}, the \textit{apparent} execution time includes the real payload activity, as well as waiting and \gls{OS}
and \gls{SW} activity. Recall that the execution times may be different~\cite{Molnar:2017:Meas,RISCVarchitecture:2017,hallaron} in the individual cases, even if the same processor executes the same instruction, but executing an instruction mix many times 
results in practically identical execution times, at least at model level. 
Note that the standard deviation of the execution times appears as a contribution to the non-parallelizable fraction,
and in this way increases the "imperfectness" of the architecture. This feature of \textit{processor}s deserves
serious consideration when utilizing a large number of processors. \textit{Over-optimizing a processor for single-thread regime hits back when using it in a many-processor environment}.

\subsection{The principle of the measurements}\label{sec:measprinciple}
When measuring performance, one faces serious difficulties, see for example~\cite{RISCVarchitecture:2017}, chapter 1,
both with making measurements and interpreting them. 
When making a measurement (i.e. running a benchmark) either on a single processor or on a system of parallelized processors,
an instruction mix is executed many times. There is, however, an crucial difference: in the second case an extra activity is also included: the job to organize the joint work. It is the reason of the 'efficiency', and it leads to critical issues in the case of extremely large number of processors.

The large number of executions averages the rather different
execution times~\cite{Molnar:2017:Meas}, with an acceptable standard deviation.
In the case when the executed instruction mix is the same, the conditions (like cache and/or memory size,
the network bandwidth, \gls{I/O} operations, etc) are different and they form the subject of the comparison.
In the case when comparing different algorithms (like results of different benchmarks),
the instruction mix itself is also different, 

Notice that the so called "algorithmic effects" -- like dealing with sparse data structures
(which affects cache behavior) or
communication between the parallelly running threads, like returning results repeatedly to the main thread
in an iteration (which greatly increases the non-parallelizable fraction in the main thread) -- 
manifest through the \gls{HW}/\gls{SW} architecture,
and they can hardly be separated.
Also notice that there are fixed-size contributions, like utilizing time measurement facilities 
or calling system services. Since $\alpha_{eff}$ is a \textit{relative} merit, the \textit{absolute}
measurement time shall be long. When utilizing efficiency data from measurements which were
dedicated to some other goal, a proper caution must be exercised with the interpretation and accuracy
of the data.

\subsection{The measurement method (bechmarking)}\label{sec:MeasBenchmarking}
Not to surprise, the method of the measurement basically affects the 
result of the measurement: the "device under test" and the "measurement device" are the same.

The benchmarks, utilized to derive numerical parameters for supercomputers, are specialized and standardized programs,
which run in the \gls{HW}/\gls{OS} environment provided by the parallelized computer under test.
One can use benchmarks for different goals. Two typical fields of utilization:
to describe  the environment the computer application runs in (a "best case" estimation), and to guess
how quickly an application will run on a given  parallelized computer (a "real-life" estimation).

If the goal is to characterize the supercomputer's \gls{HW}+\gls{OS} system itself,
a benchmark program should distort \gls{HW}+\gls{OS} contribution  as little as possible, i.e.
\gls{SW} contribution must be much lower than \gls{HW}+\gls{OS} contribution.
In the case of supercomputers, benchmark \gls{HPL} (with minor modifications) is used for this goal
since the beginning of the supercomputer age. 
The mathematical behavior of \gls{HPL} enables to 
minimize  \gls{SW} contribution, i.e.  \gls{HPL} \textit{delivers the
	possible best estimation for $\alpha_{eff}^{HW+OS}$}.

%

If the goal is to estimate the expectable behavior of an application, 
the benchmark program should imitate the structure and behavior of the application. 
In the case of supercomputers, a couple of years ago the benchmark \gls{HPCG} 
was introduced for this goal, since "\textit{\gls{HPCG} is designed to exercise computational and data access patterns that more closely match a different and broad set of important applications, and to give incentive to computer system designers to invest in capabilities that will have impact on the collective performance of these applications}"~\cite{HPCG_List:2016}. 
However, its utilization can be misleading: \textit{the ranking is only valid for the \gls{HPCG} application, and only utilizing that number of processors.}
\gls{HPCG} seems really to give better hints for designing supercomputer applications\footnote{This is why for example~\cite{DeepLearning2015} considers \gls{HPCG}  as "practical performance".}, than \gls{HPL} does.
According to our model, in the case of using the \gls{HPCG} benchmark, the \gls{SW} contribution dominates\footnote{
	Returning calculated gradients requires much more 
	sequential communication (unintended blocking).}, i.e.
\gls{HPCG} delivers the
best possible estimation for $\alpha_{eff}^{SW}$ for this class of supercomputer applications.


The different benchmarks provide different $(1-\alpha_{eff}^{SW})$ contributions to
the non-parallelizable fraction (resulting 
in different efficiencies and ranking~\cite{DifferentBenchmarks:2017}), 
so comparing results (and especially establishing ranking!)
derived using different benchmarks shall be done with maximum care.
\textit{Since the efficiency depends heavily on the number of cores (see also Fig.~\ref{fig:EffDependence2018Log} and Eq.~(\ref{eq:soverk})), the different configurations shall be compared using the same benchmark and the same number of processors (or same $R_{Peak}$).}\footnote{This is why it is misleading in some cases on the TOP500 lists to compare \gls{HPL} benchmark result, measured with all available cores, with \gls{HPCG} benchmark, measured with a small fragment of the cores.}

\section{The inherent limitations of supercomputing}
\label{sec:InherentLimit}

The limitations can be derived basically in two ways.
Either the experiences based on the implementations can be utilized to draw conclusions (as an empirical technical limit), or some theoretical assumptions can be utilized to derive a kind of theoretical  limit. The first way can be followed only if a large number of rigorously verified data are available for drawing conclusions. This method will be followed in connection with the supercomputers, where the reliable database TOP500~\cite{Top500:2016} is available. This method is absolutely empirical, and results only in an "up to now" achieved value, so one cannot be sure whether the experienced limitation is just a kind of engineering imperfectness. The other method results in a theoretical upper bound, and one cannot be sure whether it can technically be achieved. It is a strong confirmation, however, that the two ways lead to the same limitation: what can be achieved theoretically, is already achieved in the practice.

The technical implementation of the parallelized sequential 
computing systems shows up an infinite variety, so it is not really possible to describe all of them in a uniformed scheme. Instead,
some originating factors are mentioned and their corresponding term 
in the model named. At this point the simplicity of the model
is a real advantage: all possible contributions shall be classified as parallelizable or non-parallelizable ones only.
The model uses time-equivalent units, so all contributions are expressed with time, independently of their origin.

That parallel programs have inherently sequential parts (and so: inherent performance limit) is known since decades:
"\textit{Amdahl argued that most parallel programs have
some portion of their execution that is inherently serial
and must be executed by a single processor while others
remain idle.}"~\cite{ScalingParallel:1993}
Those limitations follow immediately from the physical implementation
and the computing paradigm;
it depends on the actual conditions, which of them will dominate. \textit{It is crucial to understand that the decreasing efficiency (see Equ.~(\ref{eq:soverk})) is coming
	from the computing paradigm itself rather than from some kind of engineering imperfectness.
	This inherent limitation cannot be mitigated without changing the computing/implementation principle.}

\subsection{Propagation delay PD}

In the modern high clock speed processors it is increasingly hard
to reach the right component inside the processor at the right time,
and it is even more hard, if the \gls{PU}s are at a distance
much larger than the die size. Also, the technical implementation
of the interconnection can seriously contribute.

\subsubsection{Wiring}

As discussed in \cite{LimitsOfLimits2014},
the weight of wiring compared to the processing is continuously increasing. The gates may become (much) faster,
but the speed of light is an absolute limit for the signal propagation on the wiring connecting them. 
This is increasingly true when considering large systems:
the need of cooling the modern high density processors increases the length of wiring between them.  

\subsubsection{Physical size}

Although the signals travel in a computing system with nearly
the speed of light, with increasing the physical size of the computer system a considerable time passes between issuing and receiving a signal,
causing the other party to wait, without making any payload job.
At the today's frequencies  and chip sizes a signal cannot even travel in one clock period from one side of the chip to the other, in the case of a stadium-sized supercomputer this delay can be in the order of several hundreds clock cycles.
Since the time of Amdahl, the ratio of the computing to the propagation time drastically changed, so --as~\cite{LimitsOfLimits2014} calls the attention to it--
it cannot be neglected any more, although presently it is not (yet) a major dominating term.

As long as computer components are in proximity in range $mm$,
contribution by \gls{PD} can be neglected, but the distance of \gls{PU}s in supercomputers are typically in $100~m$ range, and the nodes of a cloud system can be geographically far from each other, so considerable propagation delays can also occur.

\subsubsection{Interconnection}

The interconnection between cores happens in very different contexts, from the public Internet connection between clouds through the various connections used inside supercomputers down to the \gls{SoC} connections.
The \gls{OS} initiates only accessing the processors, after that  \gls{HW} works partly in parallel with the next action of the \gls{OS} and with other actions initiating accessing other processors.
This period is denoted in Fig.~\ref{fig:Ourmodel} by $T_x$. After the corresponding signals are generated, they must reach the target processor, that is they need some propagation time.
\gls{PD}s are denoted by $PD_{x0}$ and  $PD_{x1}$, corresponding to actions delivering input data and result, respectively.
This propagation time (which of course occurs in parallel with actions on other processors, but which is a sequential contribution within the thread) depends strongly on how the processors are interconnected: 
this contribution can be considerable if the distance to travel is large or message transfer takes a long time
(like lengthy messages, signal latency, handshaking, store and forward operations in networks, etc.).

Although sometimes even in quite large-scale systems like~\cite{SpiNNaker:2013} Ethernet-based internal communication is deployed, it is getting accepted that \textit{"The idea of using the popular shared bus to implement the communication medium [in large systems] is no longer acceptable, mainly due to its high contention."}~\cite{ReconfigurableAdaptive2016}.

\subsection{Other sources of delay}
The internal operation of the processors can also contribute to the issues experienced in the large parallel computing systems.

\subsubsection{Internal latency}
The instruction execution micro-environment can be quite different for the different parallelized sequential systems. Here it is interpreted as the internal non-payload time around executing (a bunch of) machine instructions, like waiting for the instruction being processed in the pipeline, the bus being disabled for some short period, 
copying data between address spaces,
speculating or predicting.

\subsubsection{Accelerators}

It is a trivial idea that since the single processor performance cannot be increased any more, some external computing accelerator (\gls{GPU}(s)) shall be used.
However, because of the \gls{SPA} the data must be
copied to the memory of the accelerator, and this takes time. 
This non-payload activity is a kind of sequential contribution
and surely makes the value of $(1-\alpha_{eff})$ worse.
The difference is negligible at low number of cores,
but in large-scale systems it strongly degrades the efficiency, see section~\ref{sec:UsingGPGPU}.

\subsubsection{Complexity}
The processors are optimized for single-processor performance. As a result, they attempt to make more and more operations in a single clock cycle, and doing so 
introduces a limitation for the length of the 
clock period itself:
\textit{"we believed that the ever-increasing complexity of superscalar processors would have a
	negative impact upon their clock rate, eventually leading to a leveling off of the rate of increase in microprocessor performance"}.~\cite{EPIC:2000}

\subsection{The computing paradigm}\label{sec:ComputingParadigm}

At the time when the basic operating principles of the computer were formulated, there was literally only one processor which lead naturally to using the \gls{SPA}. For today, due to development of the technology, the processor became a "free resource"~\cite{SpiNNaker:2013}.
Despite of that, mainly by inertia (the preferred incremental development) and because up to now the performance could be improved even using \gls{SPA} components (and thinking), today 
the \gls{SPA} is commonly used
when building large parallel computing systems, although the importance of "cooperative computing" is already recognized and demonstrated~\cite{CooperativeComputing2015}.
The stalling of the parallel performance may lead to the need of introducing the \gls{EMPA}~\cite{IntroducingEMPA2018}.

\subsubsection{Addressing}

One of the major drawbacks of the \gls{SPA} is what resulted in 
the components (constructed for \gls{SPA} systems).
Among others the \gls{SPA} processors use \gls{SPA} memories
through \gls{SPA} buses. This also means, that only one single addressing action can take place at a time, that is why
the need for addressing processors increases linearly
with the size of the system.
Although this issue can be mitigated by segmenting, clustering, vectoring; the basic limitation effect is present.

\subsubsection{The context switching}

All applications must use \gls{OS} services and some \gls{HW} facilities to initiate themself as well as to access other processors. Because operating system works in a different (supervisor) mode,
a considerable amount of time is required for switching context. Actually, this means virtually "another processor": a different (extended) \gls{ISA}, and a new set of processor registers.
The processor registers are very useful in single-processor optimization, but their saving and restoring considerably increases the internal latency, and what is worse, also introduces 
many otherwise unneeded memory operations.
This is usually not a really crucial contribution, but under the extreme conditions represented by supercomputers (and especially if single-port memory is used), specialized operating systems must be used~\cite{SpiNNaker:2013} or the calculation must be run in supervisor mode~\cite{FuSunwaySystem2016} or every single core most run a lightweight \gls{OS}~\cite{CooperativeComputing2015}.

\subsubsection{Synchronization}

Although not explicitly dealt with here, 
notice that the data exchange between the first thread
and the other ones also contributes to the
non-parallelizable fraction and typically uses system calls,
for details see~\cite{YavitsMulticoreAmdahl2014,CriticalSectionAmdahlEyerman:2010,SynchronizationEverything2013}.
Actually, we may have communicating serial processes, which does not improve the effective parallelism at all~\cite{CommunicatingSerialProcesses:2015}.
Some classes of applications (like artificial 
neural networks) need intensive and frequent data exchange and in addition, because of the "time grid" they need to use to coordinate the operation of the "neurons" they use, they overload the network with bursts of messages.

\begin{figure*}
	\maxsizebox{\textwidth}{!}
	{
	\begin{tabular}{cc}
		\includegraphics{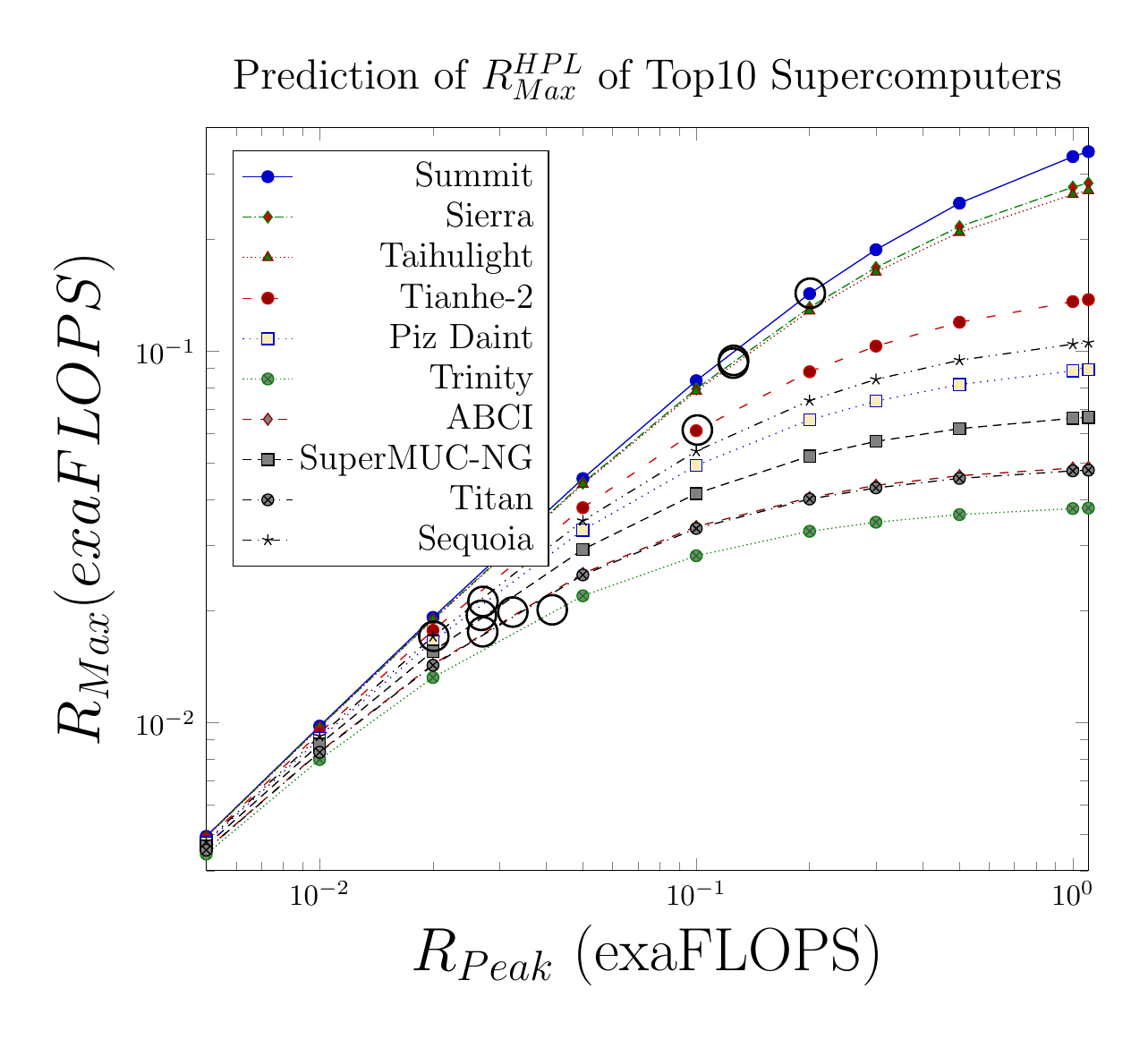}
		&
		\includegraphics{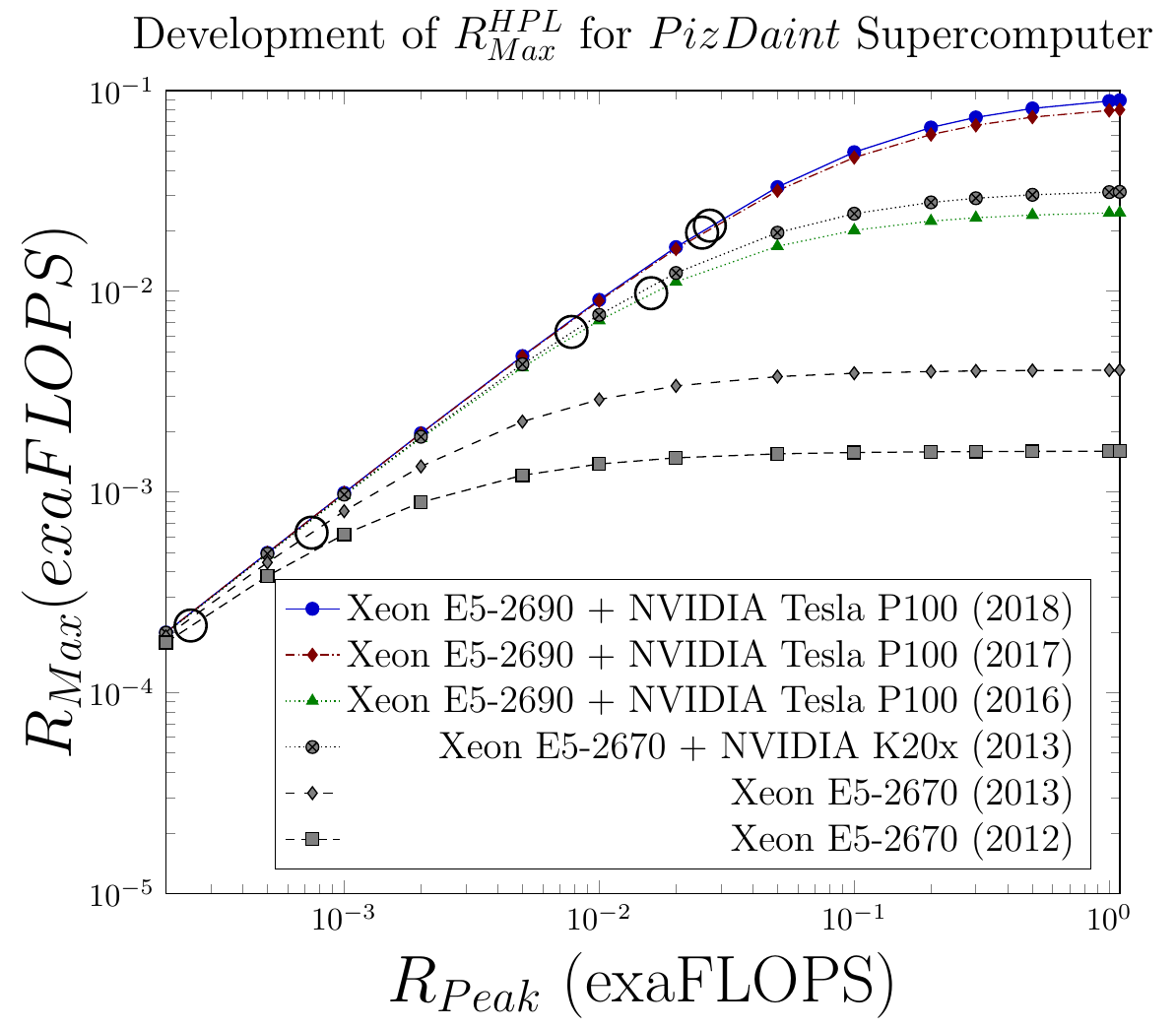}		
	\end{tabular}
    }
		\caption{a) Dependence of payload supercomputer performance on the nominal performance for the TOP10 
			supercomputers (as of November 2018) in case of utilizing the \gls{HPL} benchmark. 
			b) The timeline of the development of the payload performance $R_{Max}$ as documented in the database TOP500
			The actual positions are marked by bubbles on the diagram lines.
		}
		\label{SupercomputerTOP10Prediction}\label{fig:PizDaintRMax}	
\end{figure*}

\section{Supercomputer case studies}\label{sec:SuperCaseStudies}

From the sections above it can be concluded that 
parallelized sequential computing systems have some upper limit
on their \textit{payload performance} (the \textit{nominal performance} of course can be increased without limitation, but the efficiency decreases proportionally). In this section some case studies
on supercomputer implementations are presented, utilizing only public information. Examples of deploying the formalism on other fields of parallelized computing are given in~\cite{Vegh:2017:AlphaEff}.

\subsection{Taihulight (Sunway)}
In the parallelized  sequential computing systems implemented in \gls{SPA}~\cite{AmdahlSingleProcessor67},
the life begins in one such sequential subsystem.
In the large parallelized applications running on general purpose supercomputers,
initially and finally only one thread exists,
i.e. the minimal absolutely necessary non-parallelizable activity is to fork the
other threads and join them again.
With the present technology, no such actions can be shorter than one processor clock period\footnote{
	Taking this two clock periods as an ideal (but not realistic) case, the actual limitation will be surely (thousands of times) worse than the one calculated for this idealistic one. The actual number of clock periods depends on many factors, as discussed below.}.
That is, the absolute minimum value of the non-parallelizable fraction will be given as the ratio of the time of the two clock periods to the total execution time. 
The latter time is a free parameter in describing the efficiency, i.e. value of the effective parallelization $\alpha_{eff}$ \textit{also depends
	on the total benchmarking time} (and so does the achievable parallelization gain, too).

This dependence is of course well known for supercomputer scientists: for measuring the efficiency with better accuracy (and also for producing better $\alpha_{eff}$ values)
hours of execution times are used in practice. In the case of benchmarking 
$Taihulight$~\cite{DongarraSunwaySystem:2016}  13,298 seconds benchmark runtime was used; on the 1.45 GHz processors it means 
$2*10^{13}$ clock periods.  The inherent limit of  $(1-\alpha_{eff})$ at such benchmarking time
is $10^{-13}$ (or equivalently the achievable performance gain is $10^{13}$).
If the fork/join is executed by the \gls{OS} as usual, because of the needed context switchings $2*10^{4}$~\cite{Tsafrir:2007} clock cycles are needed rather than the 2 clock cycles considered in the idealistic case, i.e. \textit{the derived values are correspondingly by 4 orders of magnitude different; that is the performance gain cannot be above $10^9$.} 
For the development of the achieved performance gain
and the values for the top supercomputers, see Fig.~\ref{SupercomputerTimeline}.
In the following for simplicity 1.00 GHz processors (i.e. 1 ns clock cycle time) will be assumed.

The supercomputers are also distributed systems.
In a stadium-sized supercomputer a distance between the processors (cable length) about 100 m can be assumed.
The net signal round trip time is ca. 
$10^{-6}$ seconds, or $10^{3}$ clock periods, i.e. in the case of a finite-sized supercomputer the performance gain cannot be above $10^{10}$ (or $10^{6}$ if context switching also needed).
The presently available network interfaces have 100\dots200 ns latency times, and sending a message between processors 
takes time in the same order of magnitude.
Since the signal propagation time is longer than 
the latency of the network, this also means that \textit{making better interconnection 
	is not really a bottleneck in enhancing computing performance}.
This statement is underpinned also by statistical considerations~\cite{Vegh:StatisticalConsiderations:2017}.

Taking the (maybe optimistic) value $2*10^{3}$
clock periods for the signal propagation time,
the  value of the effective parallelization  $(1-\alpha_{eff})$ will be at best in the range of $10^{-10}$,
only because of the physical size of the supercomputer. 
This also means that the expectations against the absolute performance of supercomputers are excessive: assuming a 
100 Gflop/s processor and realistic physical size,
no operating system and no non-parallelizable code fraction, the achievable absolute \textit{nominal} performance (see Eq.~(\ref{eq:AmdahlMax})) is
$10^{11}$*$10^{10}~flop/s$, i.e. 1000 EFlops. To implement this, around $10^9$ processors are required. 
One can assume that the value of $(1-\alpha_{eff})$  will be\footnote{With the present technology the best achievable value is ca. $10^{-6}$, which was successfully enhanced by clustering to ca. $2*10^{-7}$ for $Summit$ and $Sierra$, and the special cooperating cores of $Taihulight$ enabled to achieve $3*10^{-8}$} around of the value $10^{-7}$.
With those very optimistic assumptions (see Equ.~\ref{eq:soverk}) the \textit{payload} 
performance for benchmark \gls{HPL} will be less than 10~Eflops, and for the real-life 
applications of class of the benchmark \gls{HPCG} it will be surely below 0.01~EFlops, i.e. lower than the payload performance of the present TOP1-3 supercomputers.

These predictions enable to assume that the presently achieved value of $(1-\alpha_{eff})$ persists also for roughly hundred times more cores. 
However, another major issue arises from the computing principle 
\gls{SPA}: on an \gls{SPA} bus only one core at a time can be addressed. As a consequence, minimum as many 
clock cycles are to be used for organizing the parallel work as many addressing steps required.
Basically, this number equals to the number of cores 
in the supercomputer, i.e. the addressing in the TOP10 positions typically needs clock cycles in the order of
$5*10^{5}$\dots$10^{7}$;
degrading the value of $(1-\alpha_{eff})$ into the range
$10^{-6}$\dots$2*10^{-5}$.
The number of the addressing steps can be mitigated
using clustering, vectoring, etc. or at the other end the processor itself can
take over the responsibility of addressing its cores~\cite{CooperativeComputing2015}.
Depending on the actual construction, the reducing factor
of clustering of those types can be in the range $10^{1}$\dots$5*10^{3}$,
i.e the resulting value of $(1-\alpha_{eff})$ is expected to be around $10^{-7}$.
Notice that utilizing "cooperative computing"~\cite{CooperativeComputing2015} enhances further
the value of $(1-\alpha_{eff})$, but it means already utilizing a (slightly) different computing paradigm: the cores have a direct connection and can communicate with the exclusion of the main memory.

An operating system must also be used, for protection and convenience. If one considers context switching with its consumed $2*10^4$ cycles~\cite{Tsafrir:2007},
the absolute limit is cca. $5*10^{-8}$, on a zero-sized supercomputer. This value is somewhat better than the limiting value derived above, but it is close to that value and surely represents a considerable contribution.
This is why $Taihulight$ runs the actual computations in kernel mode~\cite{CooperativeComputing2015}.

\subsection{Sum of the non-parallelizable contributions}

Notice the special role of the non-parallelizable activities:
independently of their origin, they are summed up as 'sequential-only'
contribution and degrade considerably the payload performance.
In systems comprising parallelized sequential processes
actions like communication (including also MPI), synchronization, accessing shared resources, etc.
\cite{CommunicatingSerialProcesses:2015,SynchronizationEverything2013,CriticalSectionAmdahlEyerman:2010,YavitsMulticoreAmdahl2014}
all contribute to the sequential-only part.
Their effect becomes more and more drastic as the number of the processors increases. 
One must take care, however, \textit{how} the communication is implemented. 
A nice example is shown in~\cite{TaihulightHPCG:2018}, how direct core to core
(in other words: direct thread to thread) communication can enhance parallelism
in large-scale systems.

\def\constProcFreq{1}	
\def\constMPE{1} 
\def\constProcPerformance{(100*\constMPE)}  
\def\constNoOfProcessors{x*1e9/\constProcPerformance} 
\def\constTotalClocks{2e13} 
\def\constContextChange{1e4} 


\def\constMicroSecToTicks{1e3*\constProcFreq}  
\def\constAlphaContext{\constContextChange/\constTotalClocks}
\def\constAlphaLoop{\constNoOfProcessors/\constTotalClocks}
\def\constAlphaOS{\constAlphaContext
	+\constAlphaLoop
}
\def\constAlphaTotal{(\constAlphaSW+\constAlphaOS)}
\def\constMinusAlpha{1-\constAlphaTotal}
\def\constEfficiency{(\constNoOfProcessors*\constAlphaTotal+\constMinusAlpha)}
\def\constRMax{x/\constEfficiency}
\def\constPropagationDelay{\constNoOfProcessors*1e-6/2e-8*2e9/\constTotalClocks*1000}



\begin{figure*}
	\maxsizebox{\textwidth}{!}
	{{
			\begin{tikzpicture}
			
			\def\constAlphaSW{2e-8}
			
			\pgfplotsset{
				xmin=0.001, xmax=1.1,
			}
			
			\begin{axis}[
			axis y line*=left,
			xlabel=$R_{Peak}(Eflop/s)$,
			ylabel=$\alpha_{eff}^{HPL}$,
			ymin=1e-10, ymax=1e-5,
			xmode=log,
			log basis x=10,
			ymode=log,
			log basis y=10,
			]
			\addplot[samples=501,domain=.001:1.1,webbrown]
			{\constAlphaSW }; \label{plot_SW}
			
			\addplot[samples=501,domain=.001:1.1,webgreen]
			{\constAlphaOS} ;\label{plot_loop}
			
			%
			\addplot[samples=501,domain=.001:1.1,webred,very thick]
			{\constAlphaTotal} ;\label{plot_total}
			
			\end{axis}
			
			\begin{axis}[
			ylabel near ticks, yticklabel pos=right,
			axis x line=none,
			ymin=0.001, ymax=1.0,
			xmode=log,
			log basis x=10,
			ymode=log,
			log basis y=10,
			ylabel=$R_{Max}^{HPL}(Eflop/s)$,
			legend style={
				cells={anchor=east},
				legend pos={south east},
			},
			]
			\addlegendimage{/pgfplots/refstyle=plot_SW}
			\addlegendentry{$\alpha^{SW}$}
			\addlegendimage{/pgfplots/refstyle=plot_loop}
			\addlegendentry{$\alpha^{OS}$}
			\addlegendimage{/pgfplots/refstyle=plot_total}
			\addlegendentry{$\alpha_{eff}$}
			%
			
			\addplot[samples=501,domain=.001:1.1,webblue,very thick] 
			{\constRMax};\label{plot_rmax}
			
			\addlegendimage{/pgfplots/refstyle=plot_rmax}
			\addlegendentry{$R_{Max}(Eflop/s)$}
			
			\end{axis}
			
			\end{tikzpicture}
		}
		{
			\begin{tikzpicture}
			
			\def\constAlphaSW{2e-6}
			
			\pgfplotsset{
				xmin=0.001, xmax=1.1,
			}
			
			\begin{axis}[
			axis y line*=left,
			xlabel=$R_{Peak}(Eflop/s)$,
			ylabel=$\alpha_{eff}^{HPCG}$,
			ymin=1e-10, ymax=1e-5,
			xmode=log,
			log basis x=10,
			ymode=log,
			log basis y=10,
			]
			\addplot[samples=501,domain=.001:1.1,webbrown]
			{\constAlphaSW }; \label{plot_SW}
			
			\addplot[samples=501,domain=.001:1.1,webgreen]
			{\constAlphaOS} ;\label{plot_loop}
			
			%
			\addplot[samples=501,domain=.001:1.1,webred,very thick]
			{\constAlphaTotal} ;\label{plot_total}
			
			\end{axis}
			
			\begin{axis}[
			ylabel near ticks, yticklabel pos=right,
			axis x line=none,
			ymin=0.001, ymax=1.0,
			xmode=log,
			log basis x=10,
			ymode=log,
			log basis y=10,
			ylabel=$R_{Max}^{HPCG}(Eflop/s)$,
			legend style={
				cells={anchor=east},
				legend pos={south east},
			},
			]
			\addlegendimage{/pgfplots/refstyle=plot_SW}
			\addlegendentry{$\alpha^{SW}$}
			\addlegendimage{/pgfplots/refstyle=plot_loop}
			\addlegendentry{$\alpha^{OS}$}
			\addlegendimage{/pgfplots/refstyle=plot_total}
			\addlegendentry{$\alpha_{eff}$}
			%
			
			\addplot[samples=501,domain=.001:1.1,webblue,very thick] 
			{\constRMax};\label{plot_rmax}
			
			\addlegendimage{/pgfplots/refstyle=plot_rmax}
			\addlegendentry{$R_{Max}(Eflop/s)$}
			
			\end{axis}
			
			\end{tikzpicture}
		}
	}
	\caption{Contributions $(1-\alpha_{eff}^X)$ to $(1-\alpha_{eff}^{total})$ and max payload performance $R_{Max}$ of a fictive supercomputer ($P=1Gflop/s$ @ $1GHz$) in function of the nominal performance.
		The blue diagram line refers to the right hand scale ($R_{Max}$ values), all others ($(1-\alpha_{eff}^{X})$ contributions) to the left scale. The left subfigure illustrates the behavior measured with benchmark \gls{HPL}. The looping contribution becomes remarkable around 0.1~Eflops, and breaks down payload performance when approaching 1~Eflops. 
		In the right subfigure the behavior measured with benchmark \gls{HPCG} is displayed. In this case the contribution of the application (thin brown line) is much higher, the looping contribution (thin green line) is the same as above. As a consequence, the achievable payload performance is lower and also the breakdown of the performance is softer.}
	\label{fig:alphacontributions}
\end{figure*}
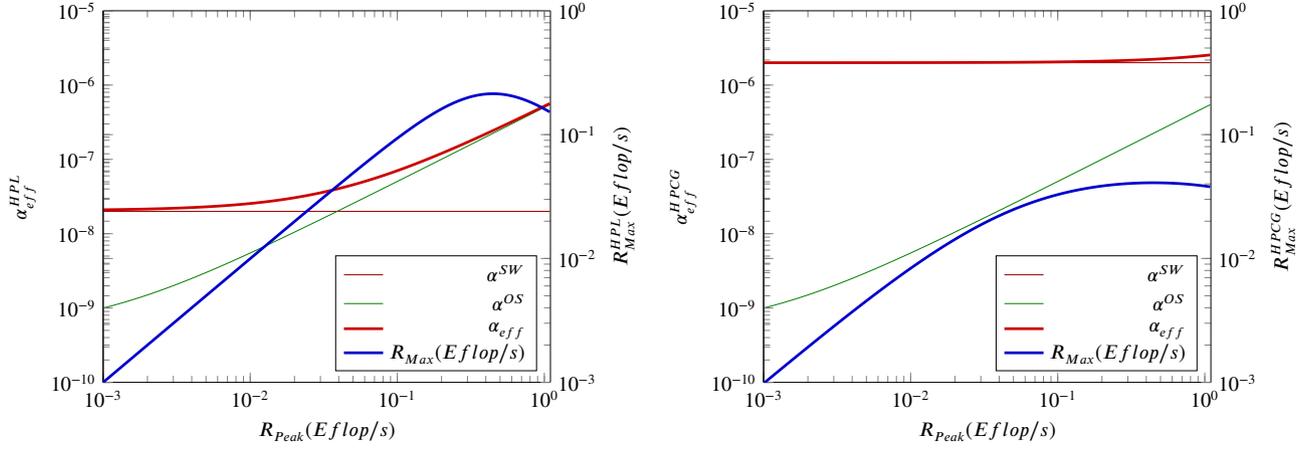

\subsection{Competition of the $\alpha_{eff}^{x}$ contributions for the dominance}

As discussed above, the different contributions of $(1-\alpha_{eff})$ depend on
different factors, so their ranking in affecting the value of $R_{Max}$
changes with the nominal performance and how the system is assembled from \gls{SPA} processors.
Fig.~\ref{fig:alphacontributions} attempts to provide a feeling on the effect of the software contribution.
A fictive supercomputer (with behavior somewhat similar to that of supercomputer $Taihulight$) is modeled. All subfigures have dual scaling. The blue diagram line refers to the right hand scale and shows the payload performance corresponding to the actual $\alpha_{eff}$ contributions; all the rest refer to the left hand scale and display $(1-\alpha_{eff}^{XX})$ (for the details see~\cite{Vegh:ExascaleComputing:2018}) contributions to the non-parallelizable fraction. The turn-back of the $(1-\alpha_{eff})$ diagram clearly shows the presence of the
"performance wall" (compare it to Fig.~1 in~\cite{ScalingParallel:1993}).

For the sake of simplicity, only those components are depicted, that have some role in forming the  $(1-\alpha_{eff})$
value.
In some other special cases other contributions may dominate. For example, as presented in~\cite{VeghBrainAmdahl:2019}, in the case of brain simulation a hidden clock signal is introduced and 
its effect is in close competition with the effect of the frequent context switchings for dominating
the achievable performance.
Notice that the performance breakdown shown in the figures were experimentally measured by~\cite{ScalingParallel:1993}, ~\cite{NeuralScaling2017}(Fig.~7)
and~\cite{NeuralNetworkPerformance:2018}(Fig.~8).

\subsection{The future of supercomputing}\label{sec:Supercomputing}
Because all of this above, in the name of the company PEZY\footnote{https://en.wikipedia.org/wiki/PEZY\_Computing: The name PEZY is an acronym derived from the greek derived Metric prefixs Peta, Eta, Zetta, Yotta
}
the last two letters are surely obsolete.
Also, no Zettaflops supercomputers will be delivered for science and military~\cite{DeBenedictis_zettaflops:2005}.

Experts expect the performance\footnote{There are some doubts about the definition of exaFLOPS, whether it means $R_{Peak}$ or $R_{Max}$, in the former case whether it includes accelerator cores, and in the latter case measured
	by which benchmark. Here the term is used as $R_{Max}^{HPL}$.} to achieve the magic 1~Eflop/s
around year 2020, see Fig.~1 in~\cite{LiaoZettaScale:2018}, 
although already question marks, mystic events and communications appeared, as the date approaches.
The authors noticed that "\textit{the performance increase of the No. 1 systems slowed down around 2013, and it was the same for the sum
performance}", but they extrapolate linearly and expect that the development continues and the "zettascale computing" (i.e $10^4$ times more than the present performance) will be achieved is just more that a decade. 
Although they address a series of important questions, \textit{the question whether building computers of such size is feasible, remains out of their sight}.

From TOP500 data, as a prediction,  $R_{Max}$ values in function of $R_{Peak}$ can be calculated, see
Fig.~\ref{SupercomputerTOP10Prediction}~a). The reported (measured) performance values are marked by bubbles on the figure.
When making that prediction, the number of processors was virtually changed for the different configurations, \textit{without correcting for the increasing looping delay}; i.e. the graphs are strongly optimistic, see also Fig.~\ref{fig:alphacontributions}.
As expected,  $R_{Max}$ values (calculated in this optimistic way)
saturate around .35~Eflop/s. Without some breakthrough in the technology and/or paradigm
even approaching the "dream limit" is not possible.

\subsection{Piz Daint}\label{sec:PizDaint}

Due to the quick development of the technology, the supercomputers have usually not many items registered in the database TOP500  on their development. One of the rare exceptions is supercomputer $Piz~Daint$. Its development history spans 6 years, two orders of magnitude in performance and used both non-accelerated computing and accelerated computing using two different accelerators.
Although usually more than one of its parameters was changed between the registered stages of its development,
it nicely underpins the statements of the paper.

Fig.~\ref{fig:PizDaintRMax}~b) displays how the \textit{payload performance} in function of the \textit{nominal performance} has developed in the case of $Piz~Daint$ (see also Fig.~2 in~\cite{VeghBrainAmdahl:2019}). The bubbles diplay the measured performance values documented in the database TOP500~\cite{Top500:2016} and the diagram lines show the (at that stage) predicted performance. As the diagram lines show the "predicted performance", the accuracy of the prediction can also be estimated through the data measured in the next stage. It is very accurate for short distances, and the jumps can be qualitatively understood with knowing the reason. In the previous section the accuracy of the predictions based on the model has been left open. This figure also validates the prediction
for the TOP10 supercomputers depicted in Fig.~\ref{SupercomputerTOP10Prediction}~a).

\begin{figure}
	\maxsizebox{1.1\columnwidth}{!}
	{
		\includegraphics[width=.8\textwidth]{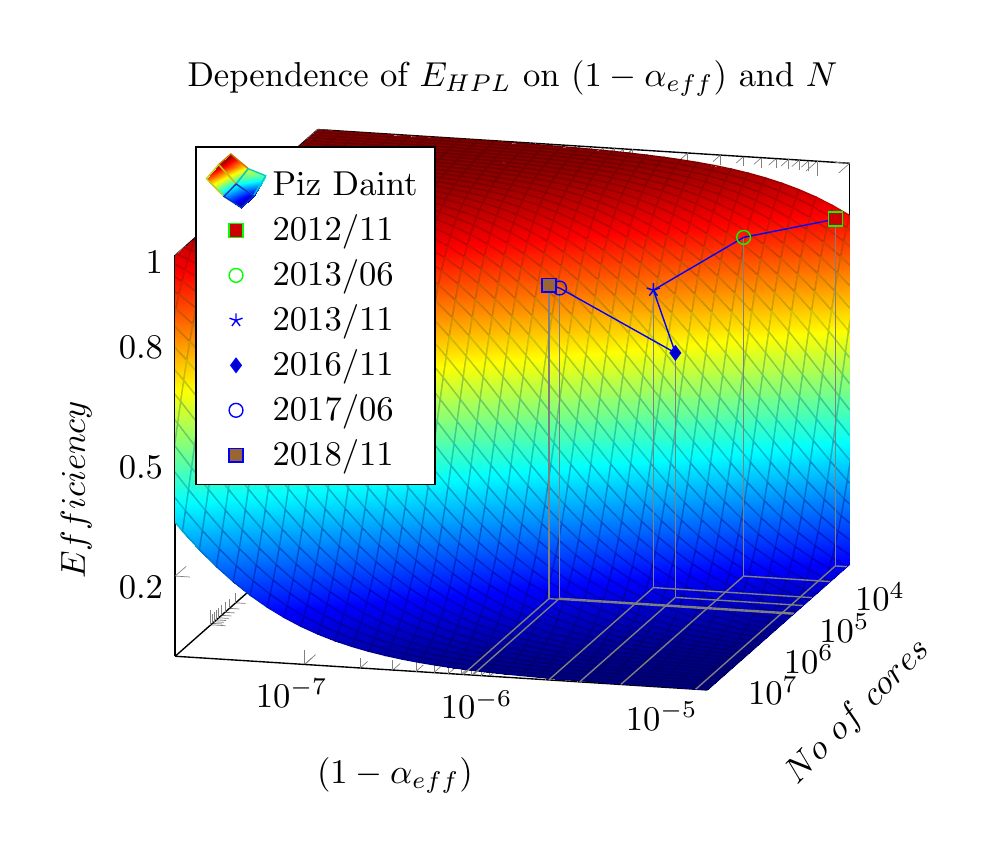}
	}
	\caption{The positions efficiency values of the supercomputer "Piz~Daint" on the two-dimensional efficiency surface in the different stages of its building; calculated from the publicly available database~\cite{Top500:2016}
	}
	\label{fig:EffDependencePizDaint}
\end{figure}

The data from the first two years of $Piz~Daint$ (non-accelerated mode of operation) 
can be compared directly. Increasing the number of the cores results in the expected higher performance, as the working point is still in the linear region of the efficiency surface. The value slightly above the predicted one can be attributed to the fine-tuning of the architecture.

Introducing accelerators resulted in a jump of payload efficiency (and also moved the working point to the slightly non-linear region, see Fig.~\ref{fig:EffDependencePizDaint}), and the payload performance 
is roughly 3 times more than it would be expected purely
on the predicted value calculated from the non-accelerated  architecture. According to the general experience~\cite{Lee:GPUvsCPU2010}, only a small fraction of the computing power hidden in the \gls{GPU} can be turned to payload performance, and the efficiency is only about 3 times higher than would be without accelerators.  

The designers might be not satisfied  with the accelerator, so they changed to another one, with a slightly higher nominal performance but much larger separated memory space. The result was disappointing: 
the slight increase of the nominal performance of the \gls{GPU} could not counterbalance
the increased time needed to copy between the separated larger address spaces, and finally resulted in a breakdown
of both the value of $(1-\alpha_{eff})$ and efficiency although the payload performance slightly increased. 
Introducing the \gls{GPU} accelerator increases the absolute performance,
but (through introducing the extra non-parallelizable component of copying the data) increases the value of $(1-\alpha_{eff})$ and decreases efficiency, for a discussion see section~\ref{sec:UsingGPGPU}.
The decrease is the more considerable the more data are to be copied. Again, the the fine-tuning has helped both
the efficiency and  $(1-\alpha_{eff})$ to have a better value.

\subsection{Gyoukou}

A nice "experimental proof" for the existence of the performance
limit is the one-time appearance of supercomputer $Gyoukou$ on the
TOP500 list in Nov. 2017. They did participate in the competition 
with using 2.5M cores (out of the 20M available) and their $(1-\alpha_{eff})$ value was $1.9*10^{-7}$, comparable with the data of $Summit$ (2.4M and $1.7*10^{-7}$). Simply, the performance bound did not enable to increase the payload performance further. 

\subsection{Brain simulation}\label{sec:BrainSimulation}
The artificial intelligence (including simulating the brain operation using computing devices) shows up exponentially growing
interest, and also the size of such systems is continuously growing.
In the case of brain simulation the "flagship goal" is to
simulate tens of billions of neurons corresponding to the capacity
of the human brain. Those definitely huge systems really go to the extremes, but also undergo the common limitation of the large-scale parallelized sequential systems. In recent studies
it was shown that using the present methods (paradigm and technology) the behavioral-level of brain is simply out of reach of  the research~\cite{NeuralNetworkPerformance:2018}.

It was shown recently~\cite{VeghBrainAmdahl:2019} that the special method
of simulating the artificial neural networks, using a "time grid",
causes a breakdown at relatively low computing performance, and that under
those special conditions the frequent context switchings and the
permanent need of synchronization are competing for dominating 
the performance of the application.

\section{Statistical underpinning}\label{sec:statistical}

For now, supercomputing has a quarter of century history and a well-documented 
and rigorously verified database~\cite{Top500:2016} on their architectural and performance data.
The huge variety of solutions and ideas does not enlighten drawing conclusions and especially making forecasts for the future of supercomputing.
The large number of available data, however, enables to draw reliable general
conclusions about some features of the parallelized sequential computing systems. Those conclusions have of course only 
statistical  validity because of the 
variety of sources of components, different technologies and ideas as well as the interplay of many factors. That is,  the result  shows up a
considerable scattering and requires an extremely careful analysis. The large number of cases, however,
enables to draw some reliable general conclusions.

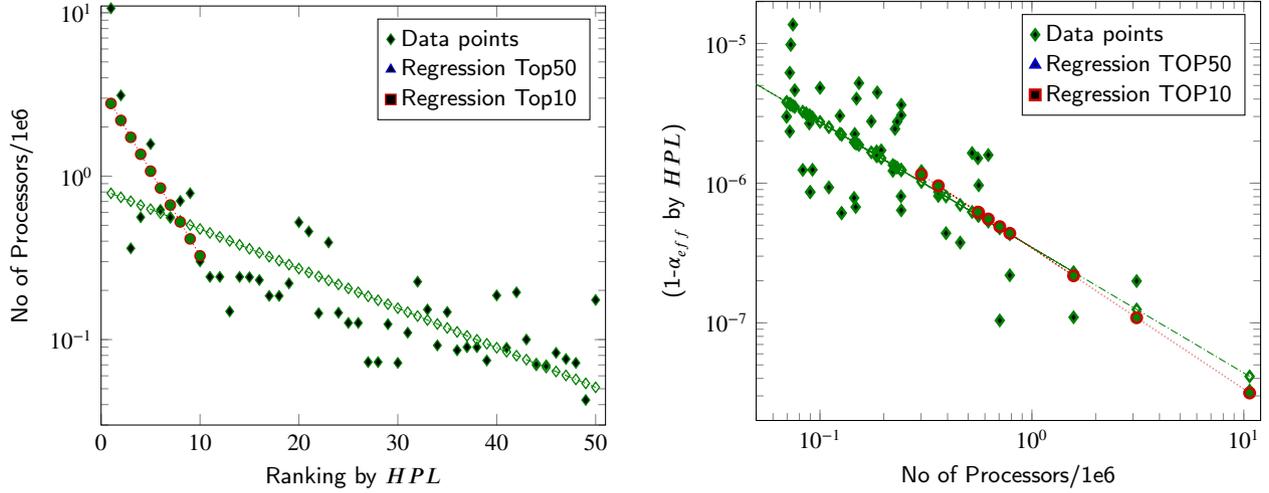
\begin{figure*}
	\maxsizebox{\textwidth}{!}
	{
		\begin{tabular}{rr}
			\tikzset{mark options={mark size=2, line width=.5pt}}
			\begin{tikzpicture}
			\begin{axis}[%
			legend style={
				cells={anchor=west},
				legend pos={north east},
			},
			cycle list name={my color list},
			xmin=0, xmax=51,
			ymin=.03, ymax=11, 
			ylabel={No of Processors/1e6},
			xlabel={Ranking  by $HPL$} ,
			ymode=log,
			scatter/classes={%
				A={ mark=diamond*,  draw=webgreen},
				N={ mark=triangle*,  draw=webblue},
				G={ mark=square*,  draw=webred}
			}
			]
			\addplot[scatter,only marks,%
			scatter src=explicit symbolic]%
			table[meta=label] {
				x y label
				1	10.649600	A
				2   3.120000    A
				3   0.361760    A
				4   0.560640    A
				5   1.572864    A
				6   0.622336    A
				7   0.556104    A
				8   0.705024    A
				9   0.786432    A
				10   0.301056    A
				11  	0.241920    A	   
				12  	0.241920    A	   
				13   0.148716    A
				14   0.241808    A
				15	0.241108    A
				16   0.231424    A
				17   0.185088    A
				18   0.185088    A
				19   0.220800    A
				20   0.522080    A
				21   0.458752    A
				22   0.144900    A
				23   0.393216    A
				24   0.145920    A
				25   0.126468    A
				26   0.126468    A
				27   0.072800    A
				28   0.072800    A
				29   0.124200    A
				30   0.072000    A
				31   0.110160    A
				32   0.225984    A
				33   0.152692    A
				34   0.092160    A
				35   0.147456    A
				36   0.086016    A
				37   0.089856    A
				38   0.089856    A
				39   0.074520    A
				40   0.186368    A
				41   0.088992    A
				42   0.194616    A
				43   0.100064    A
				44   0.069600    A
				45   0.069600    A
				46   0.082944    A
				47   0.076032    A
				48   0.072000    A
				49   0.042688    A
				50   0.174720    A
			};
			\addlegendentry{Data points}

			\addplot+[ mark=diamond,  draw=webgreen] table[y={create col/linear regression={y=Y}},
			meta=label,    /pgf/number format/read comma as period
			] {
				x Y label
				1	10.649600	A
				2   3.120000    A
				3   0.361760    A
				4   0.560640    A
				5   1.572864    A
				6   0.622336    A
				7   0.556104    A
				8   0.705024    A
				9   0.786432    A
				10   0.301056    A
				11  	0.241920    A	   
				12  	0.241920    A	   
				13   0.148716    A
				14   0.241808    A
				15	0.241108    A
				16   0.231424    A
				17   0.185088    A
				18   0.185088    A
				19   0.220800    A
				20   0.522080    A
				21   0.458752    A
				22   0.144900    A
				23   0.393216    A
				24   0.145920    A
				25   0.126468    A
				26   0.126468    A
				27   0.072800    A
				28   0.072800    A
				29   0.124200    A
				30   0.072000    A
				31   0.110160    A
				32   0.225984    A
				33   0.152692    A
				34   0.092160    A
				35   0.147456    A
				36   0.086016    A
				37   0.089856    A
				38   0.089856    A
				39   0.074520    A
				40   0.186368    A
				41   0.088992    A
				42   0.194616    A
				43   0.100064    A
				44   0.069600    A
				45   0.069600    A
				46   0.082944    A
				47   0.076032    A
				48   0.072000    A
				49   0.042688    A
				50   0.174720    A
			};
			\addlegendentry{Regression Top50}
			
			\addplot+[ mark=*,  draw=webred] table[y={create col/linear regression={y=Y}},
			meta=label,    /pgf/number format/read comma as period
			] {
				x Y label
				1	10.649600	A
				2   3.120000    A
				3   0.361760    A
				4   0.560640    A
				5   1.572864    A
				6   0.622336    A
				7   0.556104    A
				8   0.705024    A
				9   0.786432    A
				10   0.301056    A
			};
			\addlegendentry{Regression Top10}
			\end{axis}
			\end{tikzpicture}
			
			&
			\tikzset{mark options={mark size=2, line width=1pt}}
			\begin{tikzpicture}
			\begin{axis}[%
			legend style={
				cells={anchor=west},
				legend pos={north east},
			},
			cycle list name={my color list},
			xmin=0.05, xmax=12,
			ymin=2e-8, ymax=2e-5, 
			xlabel={No of Processors/1e6} ,
			ylabel={(1-$\alpha_{eff}$  by $HPL$)},
			xmode=log,
			ymode=log,
			scatter/classes={%
				A={ mark=diamond*,  draw=webgreen},
				N={ mark=triangle*,  draw=webblue},
				G={ mark=square*,  draw=webred}
			}
			]
			\addplot[scatter,only marks,%
			scatter src=explicit symbolic]%
			table[meta=label] {
				x Y label
				10.649600 3.273e-8	A
				3.120000  1.991e-7  A
				0.361760  8.094E-07  A
				0.560640  9.656E-07 A
				1.572864  1.096E-07 A
				0.622336  1.590E-06 A
				0.556104  1.507E-06  A
				0.705024  1.040E-07  A
				0.786432  2.191E-07 A
				0.301056  1.221E-06   A
				0.241920  6.399E-07 A
				0.241920  3.636E-06  A	   
				0.148716  4.028E-06  A
				0.241808  3.064E-06 A
				0.241108  8.052E-07  A
				0.231424  2.748E-06  A
				0.185088  1.689E-06 A
				0.185088  1.560E-06 A
				0.220800  1.225E-06 A
				0.522080  1.642E-06 A
				0.458752  3.756E-07 A
				0.144900  7.842E-07 A
				0.393216  4.383E-07 A
				0.145920  2.250E-06 A
				0.126468  6.107E-07 A
				0.126468  6.107E-07 A
				0.072800  9.811E-06 A
				0.072800  9.811E-06 A
				0.124200  3.036E-06 A
				0.072000  6.173E-06 A
				0.110160  9.318E-07 A
				0.225984  2.446E-06 A
				0.152692  5.204E-06 A
				0.092160  1.246E-06 A
				0.147456  6.743E-07 A
				0.086016  3.160E-06 A
				0.089856  8.635E-07 A
				0.089856  8.635E-07 A
				0.074520  1.365E-05 A
				0.186368  4.464E-06 A
				0.088992  2.677E-06 A
				0.194616  1.718E-06 A
				0.100064  4.815E-06 A
				0.069600  2.997E-06 A
				0.069600  2.997E-06 A
				0.082944  1.243E-06 A
				0.076032  4.628E-06 A
				0.072000  2.347E-06 A
				0.042688  9.587E-06 A
				0.174720  2.772E-06 A
			};
			\addlegendentry{Data points}
			
			\addplot+[ mark=diamond,  draw=webgreen] table[y={create col/linear regression={y=Y}},
			meta=label,    /pgf/number format/read comma as period
			] {
				x Y label
				10.649600 3.273e-8	A
				3.120000  1.991e-7  A
				0.361760  8.094E-07  A
				0.560640  9.656E-07 A
				1.572864  1.096E-07 A
				0.622336  1.590E-06 A
				0.556104  1.507E-06  A
				0.705024  1.040E-07  A
				0.786432  2.191E-07 A
				0.301056  1.221E-06   A
				0.241920  6.399E-07 A
				0.241920  3.636E-06  A	   
				0.148716  4.028E-06  A
				0.241808  3.064E-06 A
				0.241108  8.052E-07  A
				0.231424  2.748E-06  A
				0.185088  1.689E-06 A
				0.185088  1.560E-06 A
				0.220800  1.225E-06 A
				0.522080  1.642E-06 A
				0.458752  3.756E-07 A
				0.144900  7.842E-07 A
				0.393216  4.383E-07 A
				0.145920  2.250E-06 A
				0.126468  6.107E-07 A
				0.126468  6.107E-07 A
				0.072800  9.811E-06 A
				0.072800  9.811E-06 A
				0.124200  3.036E-06 A
				0.072000  6.173E-06 A
				0.110160  9.318E-07 A
				0.225984  2.446E-06 A
				0.152692  5.204E-06 A
				0.092160  1.246E-06 A
				0.147456  6.743E-07 A
				0.086016  3.160E-06 A
				0.089856  8.635E-07 A
				0.089856  8.635E-07 A
				0.074520  1.365E-05 A
				0.186368  4.464E-06 A
				0.088992  2.677E-06 A
				0.194616  1.718E-06 A
				0.100064  4.815E-06 A
				0.069600  2.997E-06 A
				0.069600  2.997E-06 A
				0.082944  1.243E-06 A
				0.076032  4.628E-06 A
				0.072000  2.347E-06 A
				0.042688  9.587E-06 A
				0.174720  2.772E-06 A
			};
			\addlegendentry{Regression TOP50}
			
			\addplot+[ mark=*,  draw=webred] table[y={create col/linear regression={y=Y}},
			meta=label,    /pgf/number format/read comma as period
			] {
				x Y label
				10.649600 3.273e-8	A
				3.120000  1.991e-7  A
				0.361760  8.094E-07  A
				0.560640  9.656E-07 A
				1.572864  1.096E-07 A
				0.622336  1.590E-06 A
				0.556104  1.507E-06  A
				0.705024  1.040E-07  A
				0.786432  2.191E-07 A
				0.301056  1.221E-06   A
			};
			\addlegendentry{Regression TOP10}
			
			\end{axis}
			\end{tikzpicture}
			
			\\
		\end{tabular}
	}
	\caption{a) The correlation of the TOP50 supercomputer ranking with the number of the cores
		b) The correlation of the TOP50 supercomputer ranking with the value of $(1-\alpha_{eff})$}
	\label{fig:CorrelationCores}
\end{figure*}

\subsection{Correlation between the number of cores and the achieved rank}

Since the resulting performance (and so the ranking) depends both on the number of processors  and the effective parallelization,
those quantities are correlated in Fig.~\ref{fig:CorrelationCores}.
As expected, in the TOP50 supercomputers the higher the ranking position is,
the higher is the required number of processors in the configuration,
and as outlined above, the more processors, the lower $(1-\alpha_{eff})$
is required (provided that the same efficiency is targeted).

In TOP10, the slope of the regression line on the left subfigure sharply changes
relative to the TOP50 regression line, showing the strong competition for better ranking position.
Maybe the value of the slope can provide the "cut line" between "racing supercomputers" and "commodity supercomputers".
On the right figure, TOP10 data points provide the same slope as TOP50
data points, demonstrating that to produce a reasonable efficiency,
the increasing number of cores must be accompanied with a proper decrease in value of $(1-\alpha_{eff})$, as expected from Equ.~(\ref{eq:soverk}). Furthermore, that to achieve a good ranking, a good value of $(1-\alpha_{eff})$ must be provided.
Recall that the excellent performance of $Taihulight$
shall be attributed to its special processor, deploying "Cooperative computing"~\cite{CooperativeComputing2015}.

\subsection{Deploying accelerators (GPUs)}
\label{sec:UsingGPGPU}

\begin{figure*}
	\maxsizebox{\textwidth}{!}
	{
		\begin{tabular}{rr}
			\tikzset{mark options={mark size=2, line width=.5pt}}
			\begin{tikzpicture}
			\begin{axis}[%
			legend style={
				cells={anchor=west},
				legend pos={north west},
			},
			cycle list name={my color list},
			xmin=0, xmax=51,
			ymin=00, ymax=170, 
			ylabel={Processor performance (Gflop/s)},
			xlabel={Ranking  by $HPL$} ,
			scatter/classes={%
				A={ mark=diamond*,  draw=webgreen},
				N={ mark=triangle*,  draw=webblue},
				G={ mark=square*,  draw=webred}
			}
			]
			\addplot[only marks, mark=*, draw=webblue,%
			scatter src=explicit symbolic]%
			table[meta=label] {
				x y label
				2	9.37	A
				42  9.25    A
				50  9.37    A
			};
			\addlegendentry{Accelerated}

			\addplot[only marks, mark=diamond*, draw=webgreen,%
			scatter src=explicit symbolic]%
			table[meta=label] {
				x y label
				1	11.78	N
				5  12.8   N
				6  44.8    N
				7  44.8    N
				8  16.1  N
				9  12.8  N
				10  36.8  N
				11  33.6  N
				12  52.9  N
				13  66.96 N
				14  44.8  N
				15  29.5  N
				16  41.6  N
				17  40.0  N
				18  36.8  N
				19  30.4  N
				20  18.4  N
				21  12.8  N
				22  36.8  N
				23  12.8  N
				24  36.8  N
				25  33.6  N
				26  33.6  N
				29  26.8  N
				31  31.6  N
				32  21.6  N
				34  35.2  N
				35  21.6  N
				36  41.6  N
				37  33.6  N
				38  33.6  N
				41  35.4  N
				43  36.8  N
				44  41.6  N
				45  41.6  N 
				46  31.6  N
				47  40    N
				48  35.2  N    	
			};
			\addlegendentry{Non-accelerated}
			
			\addplot[only marks, mark=triangle*, draw=webred,%
			scatter src=explicit symbolic]%
			table[meta=label] {
				x y label
				3	70.0	G
				5   48.4    G
				27   84.2    G
				28   84.2    G
				30   83.9    G
				33   36.7    G
				39   79.4    G
				40   25.2    G
				49   69.4    G
			};
			\addlegendentry{GPU-accelerated}
			
			\addplot+[ mark=o,  draw=webblue] table[y={create col/linear regression={y=Y}},
			meta=label,    /pgf/number format/read comma as period
			] {
				x Y label
				2	9.37	A
				42  9.25    A
				50  9.37    A
			};
			\addlegendentry{Regression of accelerated}
			
			\addplot+[ mark=diamond,  draw=webgreen] table[y={create col/linear regression={y=Y}},
			meta=label,    /pgf/number format/read comma as period
			] {
				x Y label
				1	11.78	N
				5  12.8   N
				6  44.8    N
				7  44.8    N
				8  16.1  N
				9  12.8  N
				10  36.8  N
				11  33.6  N
				12  52.9  N
				13  66.96 N
				14  44.8  N
				15  29.5  N
				16  41.6  N
				17  40.0  N
				18  36.8  N
				19  30.4  N
				20  18.4  N
				21  12.8  N
				22  36.8  N
				23  12.8  N
				24  36.8  N
				25  33.6  N
				26  33.6  N
				29  26.8  N
				31  31.6  N
				32  21.6  N
				34  35.2  N
				35  21.6  N
				36  41.6  N
				37  33.6  N
				38  33.6  N
				41  35.4  N
				43  36.8  N
				44  41.6  N
				45  41.6  N 
				46  31.6  N
				47  40    N
				48  35.2  N    	
			};
			\addlegendentry{Regression of nonaccelerated}
			
			\addplot+[ mark=triangle,  draw=webred] table[y={create col/linear regression={y=Y}},
			meta=label,    /pgf/number format/read comma as period
			] {
				x Y label
				3	70.0	G
				5   48.4    G
				27   84.2    G
				28   84.2    G
				30   83.9    G
				33   36.7    G
				39   79.4    G
				40   25.2    G
				49   69.4    G
			};
			\addlegendentry{Regression of GPU accelerated}
			
			\end{axis}
			\end{tikzpicture}
			
			&
			\tikzset{mark options={mark size=2, line width=.5pt}}
			
			\begin{tikzpicture}
			\begin{axis}[%
			legend style={
				cells={anchor=west},
				legend pos={north east},
			},
			cycle list name={my color list},
			xmin=0, xmax=51,
			ymin=5e4, ymax=5e9, 
			xlabel={Ranking  by $HPL$} ,
			ylabel={Performance amplification factor},
			ymode=log,
			scatter/classes={%
				A={ mark=diamond*,  draw=webgreen},
				N={ mark=triangle*,  draw=webblue},
				G={ mark=square*,  draw=webred}
			}
			]
			\addplot[only marks,mark=*, draw=webblue,%
			scatter src=explicit symbolic]%
			table[meta=label] {
				x y label
				2	0.943e7	A
				42  0.110e7    A
				50  0.676e6    A
			};
			\addlegendentry{Accelerated}
			
			\addplot[only marks, mark=diamond*, draw=webgreen,%
			scatter src=explicit symbolic]%
			table[meta=label] {
				x y label
				1	0.306e8	N
				5  0.909e7   N
				6  0.629e6    N
				7  0.662e6    N
				8  0.961e7  N
				9  0.457e7  N
				10  0.820e6  N
				11  0.156e7  N
				12  0.275e6  N
				13  0.248e6 N
				14  0.327e6  N
				15  0.124e7  N
				16  0.364e6  N
				17  0.595e6  N
				18  0.641e6  N
				19  0.820e6  N
				20  0.610e6  N
				21  0.266e7  N
				22  0.128e7  N
				23  0.228e7  N
				24  0.444e6  N
				25  0.164e7  N
				26  0.164e7  N
				29  0.329e6  N
				31  0.107e7  N
				32  0.408e6  N
				34  0.192e6  N
				35  0.8e6  N
				36  0.316e6  N
				37  0.116e7  N
				38  0.116e7  N
				41  0.373e6  N
				43  0.207e6  N
				44  0.334e6  N
				45  0.334e6  N 
				46  0.806e6  N
				47  0.216e6    N
				48  0.426e6  N    	
			};
			\addlegendentry{Non-accelerated}
			
			\addplot[only marks, mark=triangle*, draw=webred,%
			scatter src=explicit symbolic]%
			table[meta=label] {
				x y label
				3	0.124e7    G
				4   0.104e7    G
				27   0.102e6    G
				28   0.114e6    G
				30   0.162e6    G
				33   0.192e6    G
				39   0.735e5    G
				40   0.224e6    G
				49   0.104e6    G
			};
			\addlegendentry{GPU-accelerated}
			
			\addplot+[ mark=o,  draw=webblue] table[y={create col/linear regression={y=Y}},
			meta=label,    /pgf/number format/read comma as period
			] {
				x Y label
				2	0.943e7	A
				42  0.110e7    A
				50  0.676e6    A
			};
			\addlegendentry{Regression of accelerated}
			\addplot+[ mark=diamond,  draw=webgreen] table[y={create col/linear regression={y=Y}},
			meta=label,    /pgf/number format/read comma as period
			] {
				x Y label
				1	0.306e8	N
				5  0.909e7   N
				6  0.629e6    N
				7  0.662e6    N
				8  0.961e7  N
				9  0.457e7  N
				10  0.820e6  N
				11  0.156e7  N
				12  0.275e6  N
				13  0.248e6 N
				14  0.327e6  N
				15  0.124e7  N
				16  0.364e6  N
				17  0.595e6  N
				18  0.641e6  N
				19  0.820e6  N
				20  0.610e6  N
				21  0.266e7  N
				22  0.128e7  N
				23  0.228e7  N
				24  0.444e6  N
				25  0.164e7  N
				26  0.164e7  N
				29  0.329e6  N
				31  0.107e7  N
				32  0.408e6  N
				34  0.192e6  N
				35  0.8e6  N
				36  0.316e6  N
				37  0.116e7  N
				38  0.116e7  N
				41  0.373e6  N
				43  0.207e6  N
				44  0.334e6  N
				45  0.334e6  N 
				46  0.806e6  N
				47  0.216e6    N
				48  0.426e6  N    	
			};
			\addlegendentry{Regression of nonaccelerated}
			
			\addplot+[ mark=triangle,  draw=webred] table[y={create col/linear regression={y=Y}},
			meta=label,    /pgf/number format/read comma as period
			] {
				x Y label
				3	0.124e7    G
				4   0.104e7    G
				27   0.102e6    G
				28   0.114e6    G
				30   0.162e6    G
				33   0.192e6    G
				39   0.735e5    G
				40   0.224e6    G
				49   0.104e6    G
			};
			\addlegendentry{Regression of GPU accelerated}
			\end{axis}
			\end{tikzpicture}

			\\
		\end{tabular}
	}
	
	\caption{ a) The correlation of the single-processor performance of the TOP50 supercomputers (having processors with and without accelerator) with ranking, in 2017. b) The correlation of the performance gain of the TOP50 supercomputers (having processors with and without accelerator) with ranking, in 2017.}
	\label{fig:PerformanceGainVSranking}
\end{figure*}
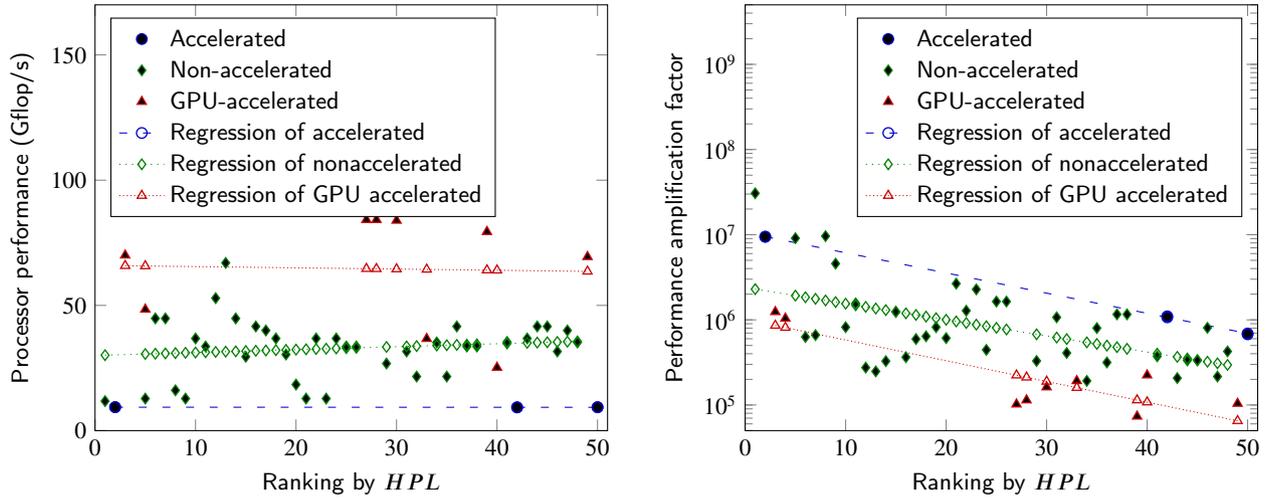

As suggested by Eq.~(\ref{eq:AmdahlMax}), the trivial way to increase
the absolute performance of a supercomputer is to increase the single-processor
performance of its processors.
Since the single processor performance has reached its limits,
some kind of accelerator (mostly \gls{GPGPU}) is frequently used for this goal. Fig.~\ref{fig:PerformanceGainVSranking} shows how utilizing accelerators influences ranking of supercomputers. 
The two important factors of supercomputers: the single-processor performance
and the parallelization efficiency
in function of ranking are displayed.

As the left side of the figure depicts,
the coprocessor accelerated cores show up the lowest performance; they really can benefit from acceleration\footnote{In the number of the total cores the number of coprocessors is included}.
The \gls{GPGPU} accelerated processors 
really increase the performance of processors by a factor of 2-3.
This result confirms results of a former study where an average factor 2.5 was found~\cite{Lee:GPUvsCPU2010}.

However, this increased performance is about 40..70 times lower than the nominal performance of the  \gls{GPGPU} accelerator. The effect is attributed to the considerable overhead~\cite{EfficiacyAPU:2011}, and it was demonstrated that with improving the transfer performance, the application performance can be considerably enhanced. 
Indirectly, that research also proved that the operating principle itself (i.e. that the data must be transferred to and from the \gls{GPU} memory; and recall that \gls{GPU}s do not have cache memory) takes some extra time. In terms of Amdahl's law,
this transfer time contributes to the non-parallelizable fraction,
i.e. increases $(1-\alpha_{eff})$, i.e. decreases the achievable performance gain.

The right side of the figure discovers this effect. The effective parallelization of the \gls{GPU} accelerated systems is nearly ten times worse than that of the coprocessor-accelerated processors and about 5 times worse than that of the the non-accelerated processors, i.e. the resulting efficiency is \textit{worse} than in the case of utilizing unaccelerated processors; this is a definite disadvance when \gls{GPU}s used in system with extremely large number of processors.

The key to this enigma is hidden in
Eq.~(\ref{eq:soverk}): the payload performance increases by a factor of nearly 3, but the value (increased by nearly an order of magnitude) of $(1-\alpha_{eff})$ is multiplied by the number of cores in the system.
In other words: while 
deploying \gls{GPGPU}-accelerated cores in
systems having a few thousand cores is advantageous, in supercomputers having processors in the range of million is a rather expensive way to make supercomputer performance \textit{worse}. 
This makes at least questionable whether it is worth to utilize \gls{GPGPU}s 
in large-scale supercomputers. 
For a discussion see section~\ref{sec:PizDaint}, for a direct experimental proof see~Figs.~\ref{fig:PizDaintRMax} and \ref{fig:EffDependencePizDaint}.

As the left figure shows, neither kind of acceleration shows correlation between the ranking of supercomputer and the type of the acceleration. 
Essentially the same is confirmed by the right side of the figure:
the performance amplification raises with the better ranking position, and the slope is higher for any kind of acceleration:
to move the data from one memory to other takes time.

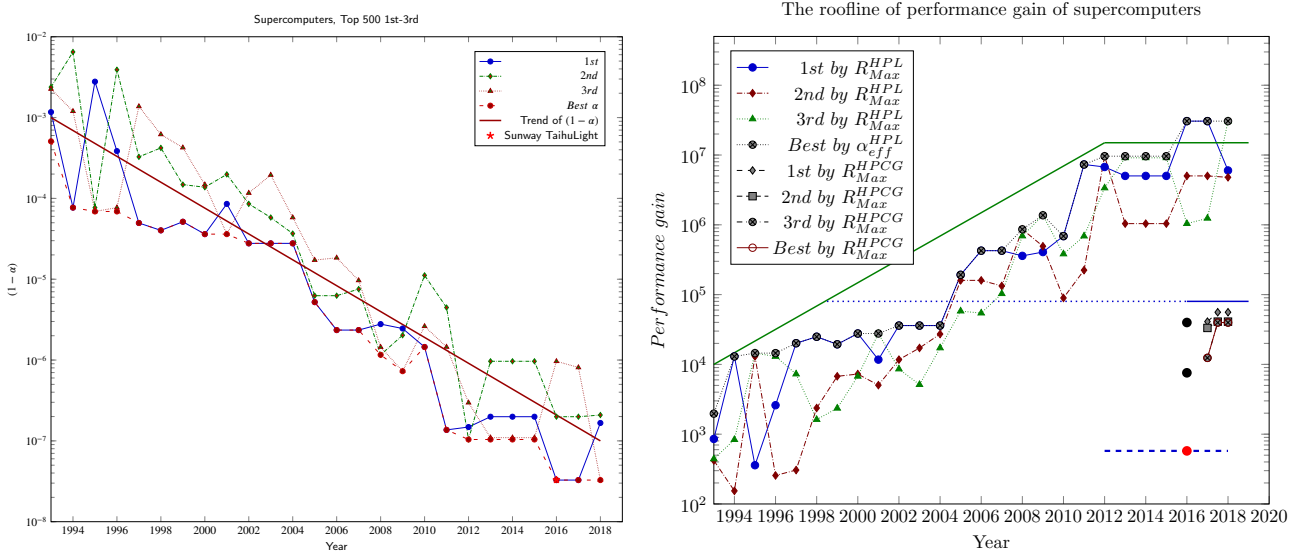
\begin{figure*}
	\maxsizebox{\textwidth}{!}
	{
		\begin{tabular}{cc}
			\begin{tikzpicture}[scale=.95]
			\begin{axis}
			[
			title={Supercomputers, Top 500 1st-3rd},
			width=\textwidth,
			cycle list name={my color list},
			legend style={
				cells={anchor=east},
				legend pos={north east},
			},
			xmin=1993, xmax=2019,
			ymin=1e-8, ymax=1e-2, 
			xlabel=Year,
			/pgf/number format/1000 sep={},
			ylabel=$(1-\alpha)$,
			ymode=log,
			log basis x=2,
			]
			\addplot table [x=a, y=b, col sep=comma] {Top500-0.csv};
			\addlegendentry{$1st $}
			\addplot table [x=a, y=c, col sep=comma] {Top500-0.csv};
			\addlegendentry{$2nd $}
			\addplot table [x=a, y=d, col sep=comma] {Top500-0.csv};
			\addlegendentry{$3rd $}
			\addplot table [x=a, y=e, col sep=comma] {Top500-0.csv};
			\addlegendentry{$Best\ \alpha$}
			\addplot[ very thick, color=webbrown] plot coordinates {
				(1993, 1e-3)  
				(2018,1e-7) 
			};
			\addlegendentry{Trend of $(1-\alpha)$}
			\addplot[only marks, color=red, mark=star,  mark size=3, very thick] plot coordinates {
				(2016,33e-9) 
			};
			\addlegendentry{Sunway TaihuLight}
			\end{axis}
			\end{tikzpicture}
			&
			\includegraphics[width=\textwidth]{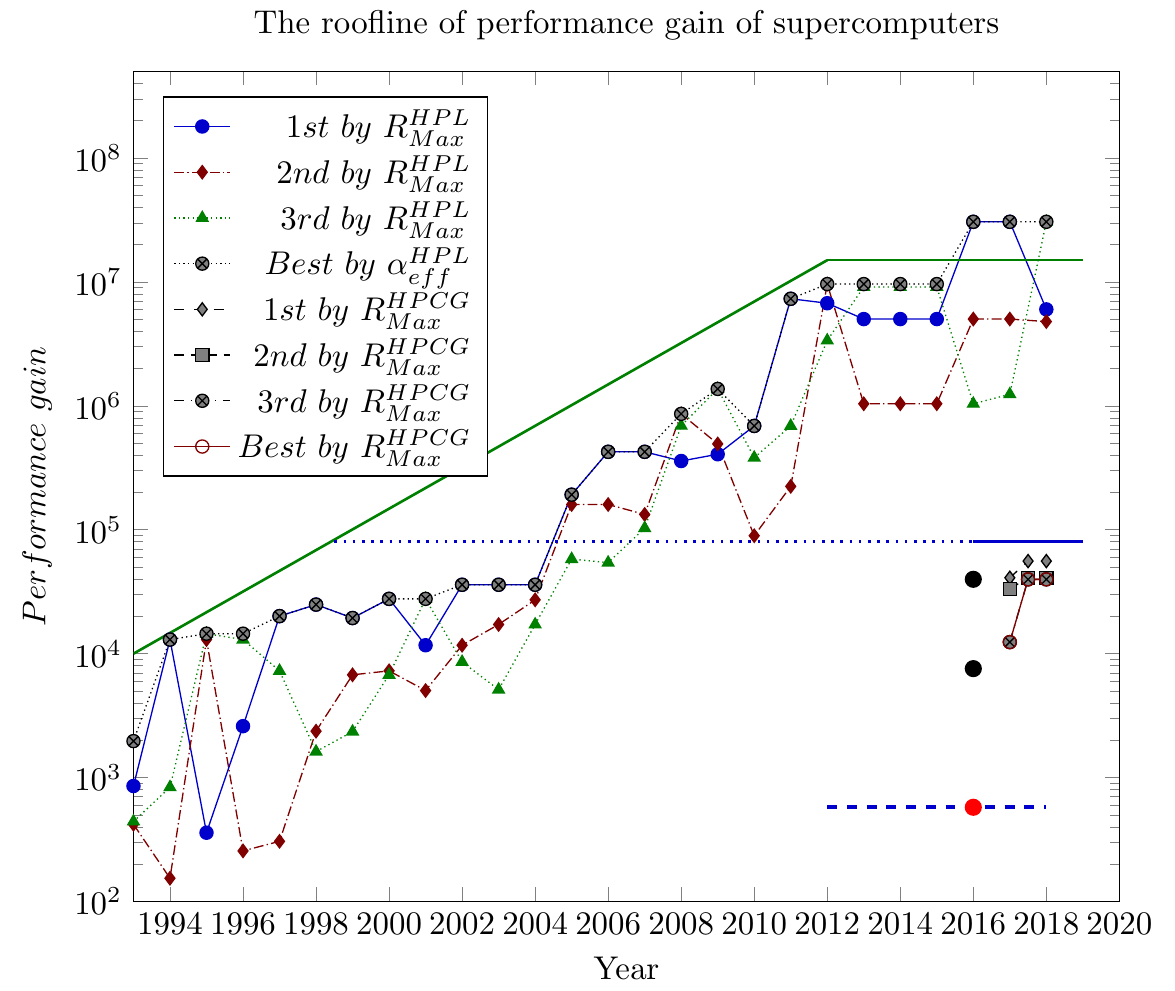}
		\end{tabular}
	}
	\caption{a) The trend of the development of ($1-\alpha$) in the past 25 years, based on the
		first three (by $R_{max}$) and the first (by ($1-\alpha$)) in the year in question.
		b)The performance gain of supercomputers modeled as "roofline"~\cite{WilliamsRoofline:2009}, as measured with the benchmarks \gls{HPL} and \gls{HPCG}, and the one concluded for brain simulation}
	\label{fig:roofline}\label{SupercomputerTimeline}
\end{figure*}

\subsection{The supercomputer timeline and the "roofline" model}
As a quick test,  Equ.~(\ref{eq:alphafromr}) can be applied to data from~\cite{Top500:2016}, see Fig.~\ref{SupercomputerTimeline}~a). 
As shown, supercomputer history is about the development of the effective parallelism,
and \textit{Amdahl's law formulated by Equ.~(\ref{eq:alphafromr}) 
	is actually what Moore's law is for the size of electronic components}.\footnote{They are also connected by that their validity terminates in the present technological epoch.}
(The effect of Moore's law is eliminated when calculating $\frac{R_{Max}}{R_{Peak}}$.) To understand the behavior of the trend line,
just recall Equ.~(\ref{eq:soverk}): to increase the absolute performance,
more processors shall be included, and to provide a reasonable efficiency,
the value of $(1-\alpha)$ must be properly reduced.

The "roofline" model~\cite{WilliamsRoofline:2009} is successful in fields where some resource 
limits the maximum performance resulting from the interplay of the other components. 
Here the limiting resource is the performance gain (originated from Amdahl's law, the technical implementation and the computing paradigm together)
that limits the engineering solutions.
Their interplay may result in more or less perfect performance gains, but no combination enables to exceed that absolute limit.

The two roofline levels displayed in Fig.~\ref{fig:roofline}~b), right side correspond to the values measurable with the benchmarks \gls{HPL} and \gls{HPCG}, respectively.
The latter benchmark has documented history in three items only, but the data are convincing enough to set a reliable roofline level. Here the contribution of the
\gls{SW} dominates ("the real-life" application class),
so the architectural solution makes no big difference, see also section~\ref{sec:MeasBenchmarking}.

The third roofline level (see section~\ref{sec:BrainSimulation}) is inferred from the single available measured data~\cite{NeuralNetworkPerformance:2018}. The two smaller black dots show the performance 
data of the full configuration (as measured by the benchmark \gls{HPCG}) of the two supercomputers the authors had access to, and the red dot denotes the saturation value they experienced (and so they did not 
deploy more hardware). This strongly supports the assumption, that the achievable supercomputer performance depends on the type of the application. The roof level concluded from the performance gain measured by benchmark \gls{HPL} only measures the effect of organizing the joint work plus the fork/join operation.

The roof level concluded from \gls{HPCG} benchmark is about two orders of magnitude lower because of the increased amount of communication needed for the iteration. In the case of the brain simulation the top level of the performance gain is two more orders of magnitude lower because of the need of more intensive communication (many other neurons must also be periodically informed  on the result of the neural calculation).
In the case of AI networks the intensity of communication is between the last two (depending
on the type and size of the network),
so correspondingly the achievable performance gain must also reside between the last two roof levels.

The figure demonstrates how the "communication-to-computation
ratio" introduced by~\cite{ScalingParallel:1993}
affects the achievable parformance gain. 
Notice that the achieved performance gain
(=speedup) of brain simulation is about $10^3$
and based on the amount of communication, 
Artificial Neural Networks also cannot show up
much higher performance gain. The bottleneck is \textit{not} the performance of the floating operations; rather the need of communication 
and the exceptionally high "communication-to-computation
ratio".
Without the need of organizing the joint work there would not be a roof level at all:
adding more cores would increase the performance gain with a permanent slope. 
With having non-zero non-parallelizable contribution the roof level appears and the 
higher that contribution is, the lower is the value of the roof level
(or, in other words: 
the higher is the non-parallelizable contribution
the lower is the nominal performance at which the roofline effect appears;
in the figure expressed in years).

The results of the benchmark \gls{HPL} show a considerable scatter and some points are even above the roofline.
This benchmark is sensitive to the architectural solutions like  clustering (internal or external), accelerators, absolute performance, etc.,
since here no \textit{ab ovo} dominating component is present.  Because of this, some "relaxation time" was and will be needed until a right combination resulting in a performance gain approaching the roofline was/will be found.

The three points above the roofline belong to the same supercomputer $Taihulight$. Its "cooperating processors"~\cite{CooperativeComputing2015} work using a slightly different computing paradigm: they slightly violate the principles of the \gls{SPA}, which is applied by all the rest of the supercomputers. Because of this, its performance gain is limited by a slightly different roofline: changing the computing paradigm (or the principles of implementation) changes the rules of the game. 
This also hints a possible way out: the computing paradigm shall be modified~\cite{RenewingComputingVegh:2018,IntroducingEMPA2018} in order to introduce higher roofline level.

\section{Summary}

The payload performance of parallelized sequential computing systems 
has been analyzed both theoretically and using the supercomputer database with well-documented performance values. It was shown that  both the (strongly simplified) theoretical description and the empirical trend show 
up limitation for the payload performance of large-scale parallelized computing, at the same value.
The difficulties experienced in building ever-larger supercomputers 
and especially utilizing artificial intelligence applications on supercomputers or building brain simulators from \gls{SPA} computer components convincingly prove that the present supercomputing has achieved what was enabled by the computing paradigm and implementation technology. To step further to the next level~\cite{ComputingPerformance:2011} a real rebooting is required,
among others renewing computing~\cite{RenewingComputingVegh:2018} and introducing a new computing paradigm~\cite{VeghEMPA:2016}. The "performance wall"~\cite{VeghPerformanceWall:2019} was hit.


\bibliography{cas-refs}

\bibliography{../../CommonBibliography}

\end{document}